\documentclass[12pt]{cernart}
\usepackage{epsfig}
\usepackage{amssymb}
\usepackage{pifont}
\usepackage{cite}
\usepackage{dcolumn}

\begin{document}
%%%%%%%%\psdraft %non mette le figure
%%%%%%%%%%%%%%%%%%%%%%%%%%%%%%% titlepage %%%%%%%%%%%%%%%%%%%%%%%%%%%%%%%%%%%%
\begin{titlepage}
  \docnum{CERN--EP/2000--026}
  \date{1 February 2000}
  \title{Uncertainties due to imperfect knowledge of systematic
         effects: \\ 
        general considerations and approximate formulae}
         \author{G. D'Agostini\Instref{xx} and M. Raso\Instref{yy}}
\Instfoot{xx}{ Universit{\`a} ``La Sapienza'' and
        Sezione INFN di Roma 1, Rome, Italy, and CERN, Geneva, Switzerland\\
        {\rm Email}: {\tt giulio.dagostini@roma1.infn.it}; 
        {\rm URL}: {\tt http://www-zeus.roma1.infn.it/$^\sim$agostini}}
\Instfoot{yy}{Sezione INFN di Roma 1, Rome, Italy (currently at 
   Parco Scientifico e Tecnologico d'Abruzzo, L'Aquila, Italy) \\
   {\rm Email}: {\tt mirko.raso@roma1.infn.it}; 
   {\rm URL}: {\tt http://\-www-\-zeus.roma1.infn.it/$^\sim$raso}}

\begin{abstract}
Starting from considerations about meaning and subsequent 
use of asymmetric uncertainty intervals of experimental results, 
we review  the issue of uncertainty propagation.
We show that, using a probabilistic approach (the so-called
Bayesian approach), all sources of uncertainty can be included
in a logically consistent way. Practical formulae for
the first moments of the probability distribution are derived 
up to second-order approximations. 

\end{abstract}
\vspace{7cm}
%\submitted{(Submitted to the American Journal of Physics)}
\end{titlepage}
%%%%%%%%%%%%%%%%%%%%%%%%%%%%%%%%%%%%%%%%%%%%%%%%%%%%%%%%%%%%%%%%%%%%%%%%%%%%%
%%%%%%%%%%%%%%%%%%%%%%%%%%%%%%%%%%%%%%%%%%%%%%%%%%%%%%%%%%%%%%%%%%%%%%%%%%%%%
% from the dvips manual: put a background `DRAFT' on the page
%\special{!userdict begin 
%/bop-hook{gsave 200 30 translate 65 rotate
%           /Times-Roman findfont 216 scalefont setfont
%           0 0 moveto 0.95 setgray (DRAFT) show grestore}def end}
%
%%%%%%%%%%%%%%%%%%%%%%%%%%%%%%%%%%%%%%%%%%%%%%%%%%%%%%%%%%%%%%%%%%%%%%%%%%%%%
\setcounter{page}{2}

\section{Introduction}
\vspace{0.2cm}
\noindent
The combination in quadrature of uncertainties due
to systematic effects has become quite standard practice
in physics.
It is also common practice to add these uncertainties 
in quadrature to those from random effects.
Usually the two kinds of 
uncertainties are given separately, 
and the systematic-effect uncertainties  
are listed individually (at least for the
most relevant ones)   
in order to show the  potential
of further measurements made with the same apparatus. 
This combination rule 
has arisen as a kind of pragmatic procedure~\cite{Maxent98}, 
in analogy to the combination of standard deviations
in probability theory, although  
cannot justifiably be termed within
`conventional' (i.e. non-Bayesian) statistics. 
The same is true for the use of the covariance matrix to
handle correlated uncertainties. 

There is less agreement when the uncertainties due to systematic
effects are asymmetric and/or they  
produce asymmetric shifts in the final quantity of interest,
due to nonlinear propagation of 
uncertainty.\footnote{As is well known, asymmetric uncertainties
arise also from random effects alone, for example in 
$\chi^2$ fits if the $\chi^2$ around the minimum is not 
symmetric. 
%Also in these circumstances -- or more generally 
%any time the $\chi^2$ function is not parabolic -- 
%the results can be misleading. 
In this paper we focus
only on asymmetries deriving from systematic
effects or non-linearity.}
As a numerical example of the latter case,
take a quantity $Y$
depending on three `influence parameters' $X_1$, $X_2$ and $X_3$,
which could be calibration constants, environment quantities
or theoretical parameters. Suppose that, for the reference
values of the $X$'s, the analysis procedure gives (in arbitrary units)
$Y=1.000\,\pm\,0.050$, where the uncertainty
associated with the result is that due to random  effects.
Consider now that by `varying reasonably
the parameters $X_i$'
(the expression is intentionally left vague for the moment) 
the following
deviations from the central values occur:
$\Delta Y_{1\pm} =\,^{+0.060}_{-0.090}$,     % gaussiana +- 1 sigma
$\Delta Y_{2\pm} =\,^{+0.098}_{-0.147}$,     % triangolare +- semidispersione
and $\Delta Y_{3\pm} =\,^{+0.104}_{-0.156}$. % uniforme  +- semidispersione
An often-used practice is to combine in quadrature separately
positive and negative deviations, obtaining the following result:
$Y = 1.00\pm 0.05\,\mbox{(stat.)}\, ^{+0.15}_{-0.23}\,\mbox{(syst.)}$.
Now we are faced with the problem that 
the result of this {\it ad hoc} 
procedure has no theoretical justification. Hence
the uncertainty content  of the statement 
(i.e. its probabilistic meaning) is unclear and, as a consequence, 
it is not obvious how to make use of
this information in further
analyses, even in the simple case in which the data points
are uncorrelated.\footnote{A different problem, although also related 
to systematic effects, is the treatment of overall uncertainty
common to all data, or to large subsets of data, in fitting. 
In fact, it is  now understood that the covariance matrix 
techniques might lead to undesirable effects on the 
results~\cite{matrix,Seibert,Smith}. We intend to review this problem
in a separate paper.}  
As a matter of fact, most people remove the asymmetry 
in further analysis of the results, 
using something equivalent to the standard deviations
 to be used in $\chi^2$ fits. 
This `standard deviation' is evaluated either   
 by taking the largest value between 
$\Delta _+$ and $\Delta _-$, or by averaging the two values 
(some use the arithmetic, others the geometric average). 
The result is that in both procedures the 
uncertainty is symmetrized and 
the result is considered as if it were described, for 
all practical purposes,  by a Gaussian
model around the published best 
estimate.\footnote{A more complicated 
`prescription' is described
by the PDG~\cite{pdg}, which we report here for the convenience 
of the reader: 
{ \it ``When experimenters quote asymmetric errors $(\delta x)^+$ and 
$(\delta x)^-$ for a measurement $x$, the error that 
we use for that measurement in making an average or a fit with 
other measurements is a continuous function of these three quantities. 
When the resultant average or fit $\overline x$ 
is less than $x-(\delta x)^-$, 
we use $(\delta x)^-$; when it is greater than $x+(\delta x)^+$, we 
use $(\delta x)^+$. In between, the error we use is a linear function 
of $x$. Since the errors we use are functions of the result, 
we iterate to get the final result.
Asymmetric output errors are determined from the input errors 
assuming a linear relation between 
the input and the output quantities.''} This rule does not seem 
to be applied by others then the PDG. 
As examples of other ad hoc procedures, see 
Refs.~\cite{pascaud,alek,offset}.} 
Our main worry is not that the combined uncertainties will be 
incorrect (we anticipate that the arithmetic average of 
$\Delta _+$ and $\Delta _-$ gives indeed the correct uncertainty 
in most cases of practical interest),
but rather that the result itself can be biased with respect to what 
one could get using consistently the best knowledge concerning
the input quantities, as  will be shown in this paper. 

The purpose of this paper is to review 
the issue of uncertainty propagation, starting from 
general considerations and deducing 
approximate formulae for practical applications. 
The issue will be analysed
in the framework of the so-called Bayesian 
inference, though Bayes' theorem will never appear in this paper.
This just means that, following 
 physics intuition~\cite{Maxent98}, 
we consider it natural to talk about 
probability of true values. As a consequence, 
for this kind of application probability 
can only have the meaning of 
degree of belief.\footnote{For a physicist's introduction 
to subjective probability and Bayesian inference  
see Ref.~\cite{YR}, or Refs. \cite{ajp}
and \cite{poisson} for short accounts.}
We will show that the rule of combining in quadrature 
symmetric uncertainties is a natural consequence of the 
probabilistic approach we follow, assuming that the uncertainties
are conceptually properly defined (although the overall result 
will not depend on their precise values). Formulae to 
take into account correlations and nonlinearity effects will
also be provided. 

The paper is structured as follows. 
First, in Section \ref{sec:prob} we illustrate briefly 
what we mean by probabilistic treatment
of measurement uncertainties and why this is
meaningful only within the Bayesian approach. 
In this approach, the most general and logically consistent
way to include all kinds of uncertainties is just
a straightforward application of probability calculus. 
However, the application of the general propagation 
formula (the derivation of which is given in Appendix A) requires 
the evaluation of integrals which can rapidly become
complicated in real-life problems. Therefore, 
approximate formulae are derived based on linear and
quadratic expansion of the output quantity on the input quantities. 
This is done in Sections \ref{sec:linear} and \ref{sec:quadratic},
respectively. Several numerical examples are given, the main ones
being discussed in Section \ref{sec:examples}. 
Section \ref{sec:parabolic} shows that our approximate formulae 
can also handle the case in which $\pm 1\,\sigma$ variations
on an input quantity produce a shift in the same direction 
on the output quantity.  
Correlations are also considered. We describe in the text only
the linear case (Section \ref{sec:linear}) and refer to Appendix B
for the more complicated formulae which take into account 
second-order effects.  
The important issue of how to model uncertainties due to 
systematic effects is discussed in Section \ref{sec:typeB}. 
For practical purposes 
the probability density function (p.d.f.) of the output quantity
can be considered approximately Gaussian also in the non-linear
cases and also if correlations are present, 
thanks to the combinatorial effects analogous to those 
which make the central limit theorem work. In case of 
important non-Gaussian contributions in the input quantities,
or strong non-linearity effects in the propagation, a detailed
evaluation of the p.d.f. is needed, usually done using Monte Carlo
methods. However, deviations from normality can be checked 
from skewness and kurtosis, and we give approximate formulae 
for these quantities in Appendix B. Finally,
we show in Appendix C how it is possible, in principle, 
to get a rough evaluation of the final p.d.f., if this does not
differ much from a Gaussian; `in principle' because we understand
that the method described in Appendix C is perhaps more an academic 
exercise than a real help to practitioners, who most likely
will find it more convenient to solve the problem by Monte Carlo
integration. Concluding remarks are given in Section 
\ref{sec:conclusions}.
 
\vspace{0.2cm}
\section{Probabilistic treatment of measurement 
uncertainties}\label{sec:prob}
\vspace{0.2cm}
\noindent
In the Bayesian approach, probability is associated with uncertainty, 
whether we are interested in the yet to be observed outcome
of a measurement, or in the numerical value of the physics 
quantity. As a result of the experiment, there will be 
a p.d.f. 
$f(\mu\,|\,\mbox{data}, I)$
associated with the  numerical
value of the true value (generically called $\mu$), 
conditioned by the observed data and by the status of 
information ($I$) concerning measurement and measurand
(the conditions $\mbox{data}$ and $I$ will usually be considered 
implicit). 
Although we maintain that the proper way of learning from  data
is to make use of Bayes' theorem, it is easy to show that
in most routine\footnote{`Routine' in the sense of 
Ref.~\cite{priors}, which applies also for most
measurements in frontier research.}
(or, at least, non critical) measurements 
the usual methods of analysis can be considered as approximations
of Bayesian inference (see also Section 2.9 of Ref.~\cite{YR}). 
When these kinds of conditions hold, also the Gaussian 
approximation is usually rather good. Therefore, hereafter
we will consider a result of the kind $\mu=\hat{\mu}\pm\sigma_r$, 
where $\sigma_r$ is only due to random effects, 
equivalent to a Gaussian
p.d.f. of $\mu$, as usually perceived by 
physicists~\cite{Maxent98,YR}, i.e.
\begin{equation}
\mu=\hat{\mu}\pm\sigma_r \ \ \Longleftrightarrow \ \ 
f(\mu_r\,|\,\mbox{data}, I) = \frac{1}{\sqrt{2\,\pi}\,\sigma_r}\,
e^{-\frac{(\mu_r-\hat{\mu})^2}{2\,\sigma_r^2}}\,,
\label{eq:equivalence}
\end{equation}
where the symbol $\mu_r$ is to remind us that only uncertainties
due to random effects are considered in Eq.~(\ref{eq:equivalence}).
On the other hand, if one sticks strictly to frequentistic 
ideas, one gets `results' which have neither a meaning of 
certain statements about true values, neither the meaning of 
probabilistic statements. As a consequence, 
it is not clear what they mean, neither how they should be
combined or propagated in a logically consistent way
(see more extended discussion in Refs.\cite{clw} and \cite{qm}). 

In the probabilistic framework in which we are moving, the uncertainties
due to systematic effects can be easily and consistently 
included (at least conceptually, although the numerical
implementation can present some technical problems). 
Indeed, there are several ways to proceed, all leading to the 
same result, though each way can be more or less intuitive
or suitable for a particular application
(see Section 2.10.3 of Ref.~\cite{YR}). The 
closest one to the spirit of probabilistic inference consists
in writing explicitly $I$ to depend on other physics quantities,
which could be calibration constants, influence parameters
(temperature, pressure, etc), theoretical quantities, and so on,
plus other pieces of general knowledge not easy to model ($I_\circ$) 
and which lead the researchers to behave in a given way and to
make reasonable assumptions in the many steps 
of the experimental work. Let us indicate by $I_\circ$ this general
knowledge. The physical quantities 
on which the result can depend will be called 
{\it influence quantities} (or parameters)
and will be indicated by $h_i$. The entire set of influence quantities
will be indicated by $\mathbf{h} = \{h_1,h_2,\ldots,h_n\}$. 
In general,   
the result (\ref{eq:equivalence}), which takes into account
only random effects, is obtained using the 
set of best estimates\footnote{Note 
that sometimes the result is meant, instead,
for a {\it nominal} set of parameters, which are not necessarily 
the ones in which the experimentalists believe mostly. 
In principle, it is possible to derive the approximate formulae
considering nominal values, as has been done by one of the 
authors in Ref.~\cite{YR}.
However, when second-order effects are
taken into account the formalism becomes more complicated and, therefore,
we prefer to start from expected values and standard deviations 
on the influence quantities. The reader should be aware that the use 
of nominal values different from expected values produce shifts 
in the results which need to be corrected~\cite{YR}, similar
to those produced by nonlinearity effects which will be discussed
later in this paper.}
 of the parameters 
($\mathbf{h}_\circ$), i.e. 
\begin{equation}
f(\mu_r\,|\,\mbox{data}, I) \equiv
f(\mu\,|\,\mbox{data},\mathbf{h}_\circ,I_\circ).
\label{eq:inf_cond}
\end{equation}
The most general inference on $\mu$ will depend, instead, 
on all possible values of $\mathbf{h}$, and the resulting 
p.d.f. will be $f(\mu\,|\,\mbox{data},I_\circ)$. 
Probability theory teaches us how to get rid of the 
uncertainty about the exact value of the influence parameters.
Describing the uncertainty about the 
influence parameters with the joint p.d.f. $f(\mathbf{h}\,|\,I_\circ)$,
we obtain that the probabilistic result which takes into
account systematic uncertainties is given by 
\begin{equation}
f(\mu\,|\,\mbox{data},I_\circ)
= \int\!f(\mu\,|\,\mbox{data},\mathbf{h},I_\circ)
  \,f(\mathbf{h}\,|\,I_\circ)\,
\mbox{d}\mathbf{h}\,.
\label{eq:inf_cond2}
\end{equation}
(We use the symbol $f(\cdot)$ for all p.d.f.'s, and 
implicitly consider the integrals done over the range of 
definition of the variables.) 
If the influence parameters are perfectly known, i.e.
$f(\mathbf{h}\,|\,I_\circ)=\Pi_i\delta(h_i-h_{\circ_i})$, we reobtain
Eq.~(\ref{eq:inf_cond}), and hence  
Eq.~(\ref{eq:equivalence}), i.e. the uncertainty 
is that due to the random effects alone. Hereafter we shall 
consider implicit the general condition $I_\circ$. 

As an example, let us consider the result of a single measurement
yielding the observed value $X=x$, and in which the most relevant 
systematic effect is a not exactly known offset $Z$, 
the uncertainty about which is
described by a Gaussian p.d.f. around zero and 
standard deviation $\sigma_z$.
We have (see Ref.~\cite{ajp} for further details): 
\begin{eqnarray}
f(\mu\,|\,x,z) &=&  
 \frac{1}{\sqrt{2\,\pi}\,\sigma_r}\,
e^{-\frac{(\mu-x-z)^2}{2\,\sigma_r^2}}\,, \\
 & &\nonumber \\
f(\mu\,|\,x) &=& \int_{-\infty}^{+\infty}\!
 \frac{1}{\sqrt{2\,\pi}\,\sigma_r}\,
e^{-\frac{(\mu-x-z)^2}{2\,\sigma_r^2}}\,
 \frac{1}{\sqrt{2\,\pi}\,\sigma_z}\,
e^{-\frac{z^2}{2\,\sigma_z^2}}\,\mbox{d}z \label{eq:intz}\\
&=& 
 \frac{1}{\sqrt{2\,\pi}\,\sqrt{\sigma_r^2+\sigma_z^2}}\,
e^{-\frac{(\mu-x)^2}{2\,(\sigma_r^2+\sigma_z^2)}}\,.
\end{eqnarray}
The p.d.f. which describes $\mu$ is still centred
around the observed value $x$, 
but with a standard deviation which is the quadratic combination 
of $\sigma_r$ and $\sigma_z$. The commonly used combination
rule is recovered, but now as a theorem with well-defined
conditions, instead of just a `prescription'.

An alternative way of including systematic effects, 
very convenient for deriving 
approximate formulae, consists 
in considering a function $g$ which relates the true 
value $\mu$  to $\mu_r$ and 
of the influence factors, i.e.
\begin{equation}
\mu=g(\mu_r,\mathbf{h})\,.
\end{equation}
Therefore the uncertainty about $\mu$ is obtained 
from the propagation of uncertainties about $\mu_r$ and 
$\mathbf{h}$ (see Appendix A):
\begin{equation}
f(\mu)=\int\!f(\mu_r,\mathbf{h})\cdot \delta(\mu-g(\mu_r,\mathbf{h}))
\,\mbox{d}\mu_r\,\mbox{d}\mathbf{h}\,. 
\label{eq:prop_delta}
\end{equation}
This formula also has a simple interpretation which 
makes it convenient for Monte Carlo 
evaluation:\footnote{For example, this is 
the basic reasoning behind the methods used 
by several authors to evaluate the p.d.f.'s 
of physical quantities, like the direct CP-violation parameter
$\epsilon^\prime/\epsilon$~\cite{epsilon_th,ciuchini}, 
or the parameters of the quark mixing  
matrix~\cite{ciuchini,parodi}:
Beliefs on the input quantities (experimental and theoretical 
quantities) are propagated into the beliefs on 
$\epsilon^\prime/\epsilon$, or on $\rho$ and $\eta$,
respectively.  The result has a clear 
probabilistic meaning and is, as we shall see, rather 
insensitive to the exact shape of the input p.d.f.'s. 
Instead, the so-called `scanning'~\cite{epsilon_th,ciuchini}
or other {\it ad hoc} procedures (see e.g. Ref.~\cite{babar})
do not have such an intuitive interpretation and 
can be misleading, especially when the 
very conservative `regions of confidence' produced 
by these methods 
are improperly called 95\% C.L. regions.} 
the infinitesimal probability element $f(\mu)\,\mbox{d}\mu$
depends on `how many' elements $\mbox{d}\mu_r\,\mbox{d}\mathbf{h}$
contribute to it, each element weighted with the p.d.f.
calculated in $\{\mu_r,\mathbf{h}\}$. 

At this point, an interesting observation is that $\mu_r$ and
$h_i$ have a symmetric role in the propagation 
of uncertainty, and therefore there is no real need to keep them
separate in the formalism. 
Therefore, following the ISO {\it Guide}~\cite{ISO},
we prefer to speak, generically, of {\it input quantities}, 
and to indicate them all by $X_i$. We indicate the {\it output 
quantity} by $Y$. In many problems of interest
also the output quantities might be more than one. Their 
values are evaluated using the common data or (which is 
conceptually equivalent) making use of the same instrumentation.
In such a case we have to consider correlations among 
the output quantities even if the input quantities were
uncorrelated. Hereafter we will indicate the generic 
functions $g_i(\mathbf{X})$ with the same
symbol as the output variables, and speak about $Y_i=Y_i(\mathbf{X})$.

\vspace{0.2cm}
\section{Linear expansion around $\mbox{E}[\mathbf{X}]$,
role of central limit theorem and numerical 
implementation of the linear propagation}\label{sec:linear} 
\vspace{0.2cm}
\noindent
Having illustrated the general solution to the problem, 
it is now interesting to obtain approximate formulae 
which,  in many practical cases,
 save us from making complicated integrals.
The case in which the dependence of $Y_j$ 
on $X_i$ is approximately linear in a range of several 
 standard deviations around their expected value is well known,
and leads to the standard propagation formula of variances
and covariances. 
What is less well known is that the use of these formulae is 
justified only if the numerical values of the physics quantities 
are associated with random numbers (or uncertain numbers), 
and the probability is meant as degree of 
belief~\cite{YR,ajp}. 

The first-order expansion of $Y_i(\mathbf X)$ around the expected
values of $X_i$ gives
\begin{eqnarray}
Y_j &\approx& Y_j(\mbox{E}[\mathbf{X}])
          + \sum_i\left.
                 \frac{\partial Y_j}
                      {\partial X_i}
                 \right|_{\mbox{\small E}[\mathbf{X}]}
                             (X_i-\mbox{E}[X_i]) \label{eq:linear1} \\
  &\approx& k
          + \sum_i\left.\frac{\partial Y_j}{\partial X_i}
                  \right|_{\mbox{\small E}[\mathbf{X}]} X_i\,,
\label{eq:linear2}
\end{eqnarray}
where $\mbox{E}[\cdot]$ stands for expected value 
and the derivatives are evaluated for 
$\mathbf{x}=\mbox{E}[\mathbf{X}]$
(this will be implicit hereafter).
The second formula is very convenient to calculate the variance, having
put in $k$
all terms which do not contain $X_i$. 
Evaluating the expected values from Eq.~(\ref{eq:linear1}), 
and variances and covariances from Eq.~(\ref{eq:linear2}), we get
 (we have replaced the symbol `$\approx$'
by `$=$' to indicate that there are no further 
approximations than linearization): 
\begin{eqnarray}
\mbox{E}[Y_j] &=& Y_j(\mbox{E}[\mathbf{X}]) \,,
\label{eq:EY} \\
\sigma^2(Y_j) &=& \sum_i\left(\frac{\partial Y_j}
     {\partial X_i}\right)^2\sigma^2_i +
\left\{2\,\sum_{l<m}
\left(\frac{\partial Y_j}
     {\partial X_l}\right)
\left(\frac{\partial Y_j}
     {\partial X_m}\right)\,\rho_{lm}\,\sigma_l\,\sigma_m
\right\}\,, 
\label{eq:varY} \\
\mbox{Cov}(Y_j,Y_k) &=&
\sum_i\left(\frac{\partial Y_j}
     {\partial X_i}\right)
\left(\frac{\partial Y_k}
 {\partial X_i}\right)\sigma^2_i +
\left\{2\,\sum_{l<m}
\left(\frac{\partial Y_j}
     {\partial X_l}\right)
\left(\frac{\partial Y_k}
     {\partial X_m}\right)\,\rho_{lm}\,\sigma_l\,\sigma_m
\right\} \,, 
\label{eq:covY}
\end{eqnarray}
where $\sigma_i$ are shorthand for $\sigma(X_i)$
and $\rho_{lm}$ are the correlation coefficients, such 
that $\rho_{lm}\,\sigma_l\,\sigma_m = \mbox{Cov}(X_l,X_m)$.
The terms within $\{\cdot\}$ vanish if the input quantities 
are uncorrelated, as it often the case when relevant systematic
effects are considered. However, sometimes this is not
the case, as when several calibration constants are simultaneously 
obtained from a fit.  
Equations (\ref{eq:varY})--(\ref{eq:covY}) can be written in the
more compact form of covariance matrix transformation. However,
for the purpose of this paper, we prefer not to use the matrix
formalism, in order to separate the contributions due 
to variances and covariances. 

Equations (\ref{eq:EY})--(\ref{eq:covY}) give only some limited 
information about the joint p.d.f. of $\mathbf{Y}$, namely only 
1st and 2nd moments. 
However the central limit theorem plays the important role
of making the   p.d.f. of each $Y_j$ practically
Gaussian in most of the cases of interest (see e.g. 
examples and words of caution in Ref.~\cite{YR},
and discussion in Ref.~\cite{Higgs}). 
The joint p.d.f. can be considered for
practical purposes a multivariate Gaussian. 
Anyway, in case of doubt, it is  good practice 
to check the shape of each marginal p.d.f.
(see Appendix C).

In complex real-life cases  the derivatives are not 
performed analytically. Instead, the effects 
of the input values on the output values are evaluated 
 numerically, often by Monte Carlo techniques. 
In these cases the derivatives
can be estimated numerically by $\pm\,1\,\sigma$ variations
around the expected values. 
Calling $\Delta _{\pm_{ji}}$\footnote{We have used 
the following notation: 
$\Delta _+ = Y(\mbox{E}[X]+\sigma_X) - Y(\mbox{E}[X])$ 
and $\Delta _- = Y(\mbox{E}[X])-Y(\mbox{E}[X]-\sigma_X) $. 
Therefore, for monotonic functions around $\mbox{E}[X]$ the 
increments  
$\Delta _+$ and $\Delta _-$ have the same sign.\label{fn:delta}}
the variation
of $Y_j$ due to a variation of $X_i$ of $\pm 1\,\sigma_i$ around
$\mbox{E}[X_i]$, linearity implies that
 \begin{equation}
\frac{\partial Y_j}
           {\partial X_i}
    \approx \frac{\Delta _{+_{ji}}}{\sigma_i}
    \approx \frac{\Delta _{-_{ji}}}{\sigma_i}. 
\end{equation}
Since in the linear approximation 
$\Delta _{+_{ji}}$ and $\Delta _{-_{ji}}$ are practically equal, 
we call $\Delta _{ji}$ either of them (taking the average of the two 
if there are  small differences; the case of large differences,
hint of non-linear effects, will be discussed below). 
We get, finally, the following practical formulae for the elements
of the covariance matrix:
\begin{eqnarray}
\sigma^2(Y_j) &=& \sum_i \Delta _{ji}^2  +
\left\{2\,\sum_{l<m}
\rho_{lm}\, \Delta _{jl}\,\Delta _{jm} \right\}\,, 
\label{eq:varY1} \\
\mbox{Cov}(Y_j,Y_k) &=&
\sum_i \Delta _{ji}\,\Delta _{ki} +
\left\{2\,\sum_{l<m}
\rho_{lm}\,\Delta _{jl}\,\Delta _{km} \right\} \,.
\label{eq:covY1}
\end{eqnarray}
In the simple case of independent input quantities, 
Eqs. (\ref{eq:varY1})--(\ref{eq:covY1}) reduce to
\begin{eqnarray}
\sigma^2(Y_j) &=& \sum_i \Delta _{ji}^2  
\label{eq:varY2} \\
\mbox{Cov}(Y_j,Y_k) &=& \sum_i \Delta _{ji}\,\Delta _{ki} \left[=
\sum_i \mbox{Cov}_i(Y_j,Y_k) = \sum_i s_{ijk}\,
|\Delta _{ji}|\,|\Delta _{ki}|\right] \,,
\label{eq:covY2}
\end{eqnarray}
where $\mbox{Cov}_i(Y_j,Y_k)$ stands for the contribution 
to the covariance from the $i$th input quantity, and 
$s_{ijk}$ indicate the product of the signs of the absolute 
increments of $Y_j$ and $Y_k$ for a variation of $X_i$
($|\Delta _{ji}|$ have the meaning of standard uncertainty 
of $Y_j$ due to $X_i$ alone). 

At this point, we have to remember that $\mu_r$ defined in 
Sec.~\ref{sec:prob} is considered as one of the input quantities, 
and that in the most general case there will be many $\mu_{r_j}$,
each associated with one and only one output quantity $Y_j$. 
The resulting covariance matrix will be equal to the sum 
of the covariance matrix of the $\mu_{r_j}$ 
(they can be correlated as they 
could come from fitting procedures, unfolding, or other 
statistical techniques) and the covariance matrix due to
the systematic effects. 
Let us write down, as an easy and practical example, the formulae
for the case when we have $N$ values  $\mu_{r_j}$ and 
the influence quantities are uncorrelated: 
\begin{eqnarray}
\sigma^2(Y_j) &=& \sigma_{r_j}^2 +\sum_{i>N} \Delta _{ji}^2\,,  
\label{eq:varY2r} \\
\mbox{Cov}(Y_j,Y_k) &=& \mbox{Cov}(\mu_{r_j},\mu_{r_j}) 
%\sum_i \mbox{Cov}_i(Y_j,Y_k)\,,
 + \sum_{i>N} s_{ijk}\, |\Delta _{ji}|\,|\Delta _{ki}| \,,
% + \sum_{i>N} \Delta _{ji}\,\Delta _{ki} \,,
\label{eq:covY2r}
\end{eqnarray}
where we have taken into account that the $\Delta _{ji}$ associated 
with $\mu_{r_i}$ are given by   $\Delta _{ji}=\sigma_i\,\delta_{ij}$, 
where $\delta_{ij}$  is Kronecker symbol. In fact,  
the derivatives of $Y_j$ with respect to $\mu_{r_i}$, evaluated at 
the point of best estimate of $\mathbf{X}$,  are equal to
1 if $i=j$, and equal to 0 otherwise. 

\vspace{0.1cm}
\section{Modelling the uncertainty due to systematic effects: 
ISO type-B uncertainties}\label{sec:typeB}
\vspace{0.1cm}
\noindent
At this point it is important to define somewhat better 
how the several ingredients appearing in the previous formulae
should be evaluated. 
In fact, the results of the above formulae
have a defined probabilistic meaning only if the 
various $\Delta$'s 
are obtained as variations of 
the output quantities for  $1\,\sigma$ variations of the input
quantities, and not, generically, as reasonable variations, 
or, prudentially, as `conservative variations'. 
Now we are confronted with the problem that in the
evaluation of uncertainties due to imperfect knowledge
of systematic effects, the case in which 
the input uncertainties are evaluated from standard statistical
procedures which provide standard deviations in
an automatic way is rare. These latter 
cases would be those in which we feel comfortable. 
More often, {\it ``for estimate $x_i$ of an input quantity $X_i$ 
that has not been obtained from repeated observations,
the \ldots standard uncertainty \ldots  is evaluated 
by scientific judgement based on all the available information on the 
possible variability of $X_i$. The pool of information may include:
previous data; experience with or general knowledge of the behaviour 
and properties of relevant materials and instruments; 
manufacturer's specifications; 
data provided in calibration and other certificates; uncertainties 
assigned to reference data taken from handbooks.''}
(ISO Guide~\cite{ISO}).
This is along the spirit that {\it ``the evaluation of uncertainty is 
neither a routine task nor a purely mathematical one; it depends 
on detailed knowledge of the nature of the measurand and of the 
measurement''}~\cite{ISO}. 

\begin{figure}[!t]
\begin{center}
\begin{tabular}{|c|c|} \hline 
  & \\
\multicolumn{1}{|l|}{\small 
$\begin{array}{lcl} \mbox{E}[X] &=& x_0 [= (a+b)/2] \\
  \sigma(X) &=& \Delta x/\sqrt{3} \\
  {\cal S} &=& 0 \\
  {\cal K} &=& 9/5 \end{array}$ } & 
\multicolumn{1}{|l|}{\small 
$\begin{array}{lcl} \mbox{E}[X] &=& x_0 = (a+b)/2 \\
  \sigma(X) &=& \Delta x/\sqrt{6} \\
  {\cal S} &=& 0 \\
  {\cal K} &=& 12/5 \end{array}$ } \\
\epsfig{file=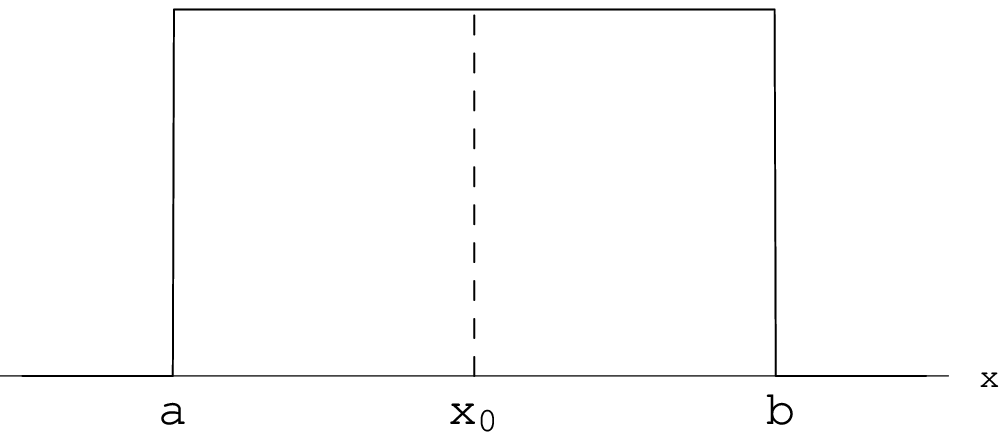,width=0.45\linewidth,clip=} &
\epsfig{file=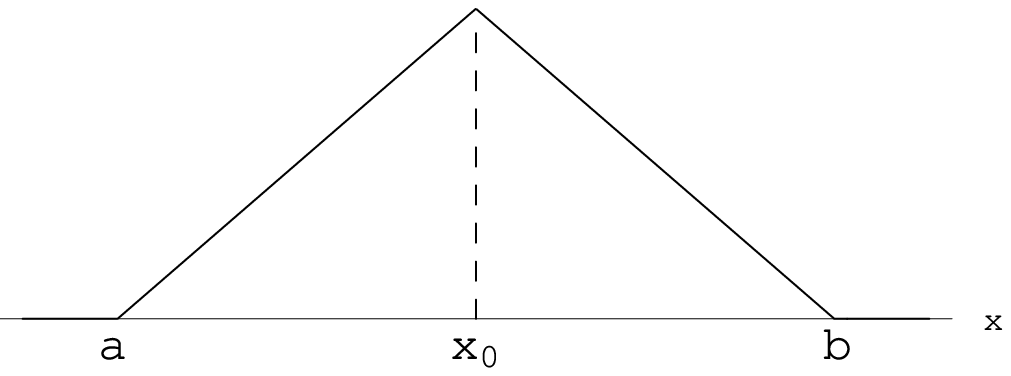,width=0.45\linewidth,height=4.2cm,clip=}\\ 
\hline
& \\
\multicolumn{1}{|l|}{\small 
$\begin{array}{lcl} 
\mbox{E}[X] &=& x_0 +(\Delta x_+ - \Delta x_-)/3 \\
\sigma^2(X) &=& (\Delta^2x_+ + \Delta^2x_- + \Delta x_+ \Delta x_-)/18 \\
{\cal S}    &=&  [3(\Delta^2x_+\Delta x_- - \Delta^2x_- \Delta x_+) \,+ \\
 && 2(\Delta^3 x_+ -\Delta^3 x_-)]\, /\,270\,\sigma^3(X) \\
%(1/(270\,\sigma^3(X)))\cdot [2(\Delta^3x_+ -\Delta^3x_-) \\
%  && + 3(\Delta^2x_+\Delta x_- - 
%  \Delta^2x_- \Delta x_+)] \\
{\cal K}    &=& 12/5 
\end{array}$ }  &
\multicolumn{1}{|l|}{\small $ \begin{array}{lcl} 
                       \mbox{E}[X] &=& x_0  \\
                       \sigma(X) &=& \sigma \\
                       {\cal S} &=& 0 \\
                       && \\
                       {\cal K} &=& 3 \end{array}$ } \\
\epsfig{file=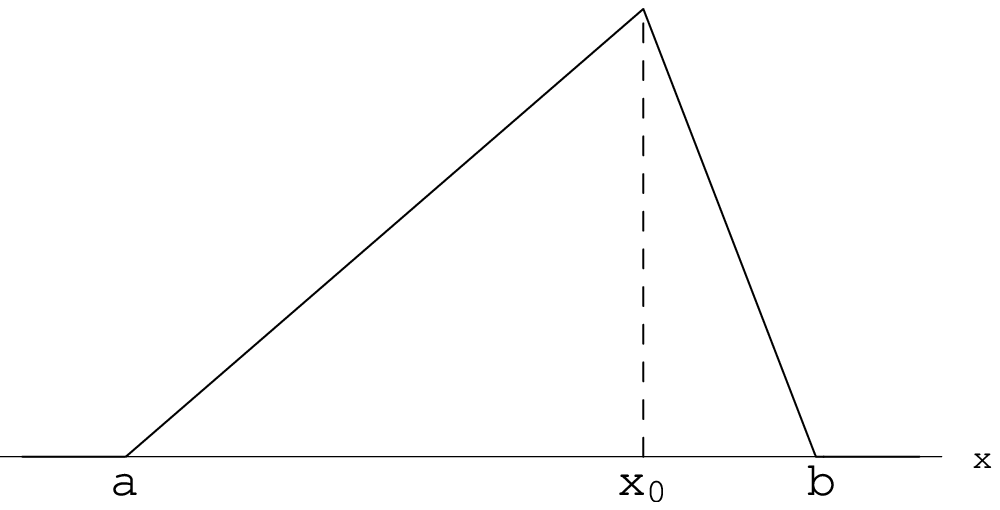,width=0.45\linewidth,height=4.2cm,
clip=} &
\epsfig{file=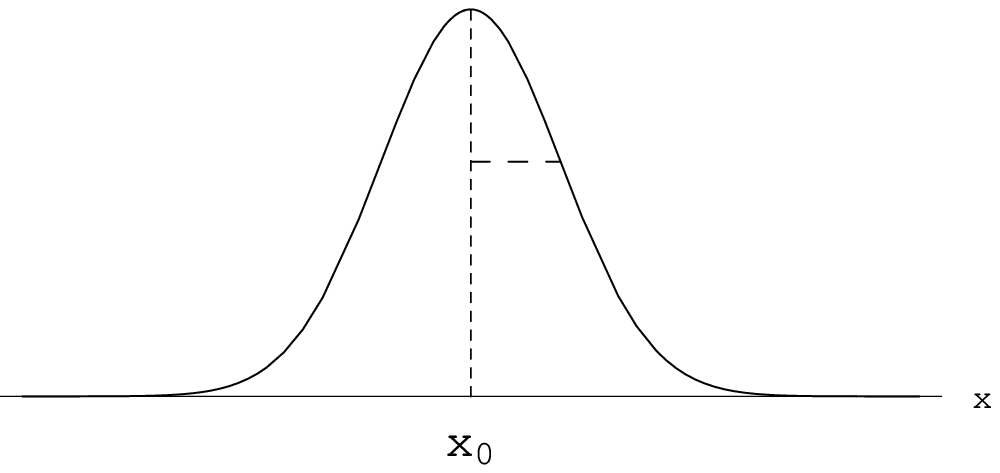,width=0.45\linewidth,height=4.2cm,clip=} \\
\hline
\end{tabular}
\end{center}
\vspace{-8.9cm} 
\hspace{+4.6cm} $\Delta x $ \hspace{+7.0cm}  $\Delta x $ 

\vspace{+5.8cm}
\hspace{+12.cm}  $\sigma $ 

\vspace{+0.5cm}
\hspace{+3.7cm}$\Delta x_- $ \hspace{+0.55cm} $\Delta x_+$ 

\vspace{+1.1cm}
\caption{\small Typical models 
to assess type-B uncertainties: 
uniform distribution,  
 symmetric triangular distribution,  
asymmetric triangular distribution, and 
Gaussian distribution. The expressions of the most relevant 
statistical parameters are reported (${\cal S}$ stands for skewness,
${\cal K}$ for kurtosis).}
\label{fig:pdfsys}
\end{figure}

Following the recommendation 
of the BIPM recommendations~\cite{BIPM},
the ISO Guide calls this kind 
of uncertainty type B,
in contrast to type-A uncertainties obtained, to say it shortly,
 by the dispersion of 
readings (see also Section 6.1.2 of Ref.~\cite{YR}). 
The evaluation of type-B uncertainties 
implies the adoption of the {\it ``viewpoint \ldots that probability 
is a measure of the degree of belief that an event will 
occur''}~
\cite{ISO}. In practice it 
requires a realistic and 
honest modelling of the case. 
The most common models are discussed 
in the ISO Guide itself:
For example, if one is practically sure that an input value
is in a certain interval, and all values inside the interval
appear similarly likely, the proper model for the uncertainty 
is a  uniform distribution. Other times the edges 
of the interval seem still to be really extreme values for the quantity;
but one tends to believe more in central values, and the 
belief decreases roughly linearly from the centre to the edges.
In this case, a more suitable distribution is a  symmetric
triangular distribution. Alternatively, if the 
belief decreases towards the edges, but the maximum belief
does not coincide with (approximately) the centre of the 
interval, it is preferable to use an asymmetric triangular
distribution. Finally, if the interval seems simply to be just
a highly probable one (e.g. 90\%, 95\% or 99\%), but also far away values  
are believed to be possible, one can use a Gaussian model 
with a standard deviation which is a suitable fraction of the 
uncertainty interval. 
Figure ~\ref{fig:pdfsys} shows 
the most common models 
to handle type-B uncertainties, together with their most
interesting statistical parameters. 

Once the mathematical model has been 
chosen, the p.d.f. of the output quantity can be evaluated
using Eq.~(\ref{eq:prop_delta}), or their first moments 
can be obtained using approximated formulae.  
It is important to realize that neither 
the choice of model, nor the value of the standard deviation
are very critical, as discussed in Refs.~\cite{YR} and ~\cite{ISO}
(see also Section 4.6 of Ref.~\cite{Higgs}). In fact, the 
central limit theorem makes the result Gaussian independently 
of the initial distribution, if none of the non-Gaussian components
dominates. Moreover, some unavoidable over- or under-estimates,
 compensate, if one makes the effort of assessing 
model and standard deviation in an honest way. Finally, 
in agreement with the ISO Guide, we think that it is 
bad practice to overestimate intentionally  
type-B uncertainties
(see Section 10.2 of Ref.~\cite{YR}). 

\begin{figure}[!h]
\begin{center}
\begin{tabular}{|ccccc|}\hline
&&&& \\
\multicolumn{1}{|l}{\small $\begin{array}{lcl} \mbox{E}[X] &=& 0.17 \\
 \sigma(X) & = & 0.42 \\
\mbox{mode} &=& 0.5 \\
\mbox{median} &=& 0.23 
\end{array}$} & & 
\multicolumn{1}{l}{\small $\begin{array}{lcl} \mbox{E}[X] &=& 0.17 \\
 \sigma(X) & = & 0.42 \\
\mbox{mode} &=& 0.5 \\
\mbox{median} &=& 0.23 
\end{array}$} & &
\multicolumn{1}{l|}{\small $\begin{array}{lcl} \mbox{E}[X] &=& 0.34 \\
 \sigma(X) & = & 0.59 \\
\mbox{mode} &=& 0.45  \\
\mbox{median} &=& 0.37 
\end{array}$} \\
\epsfig{file=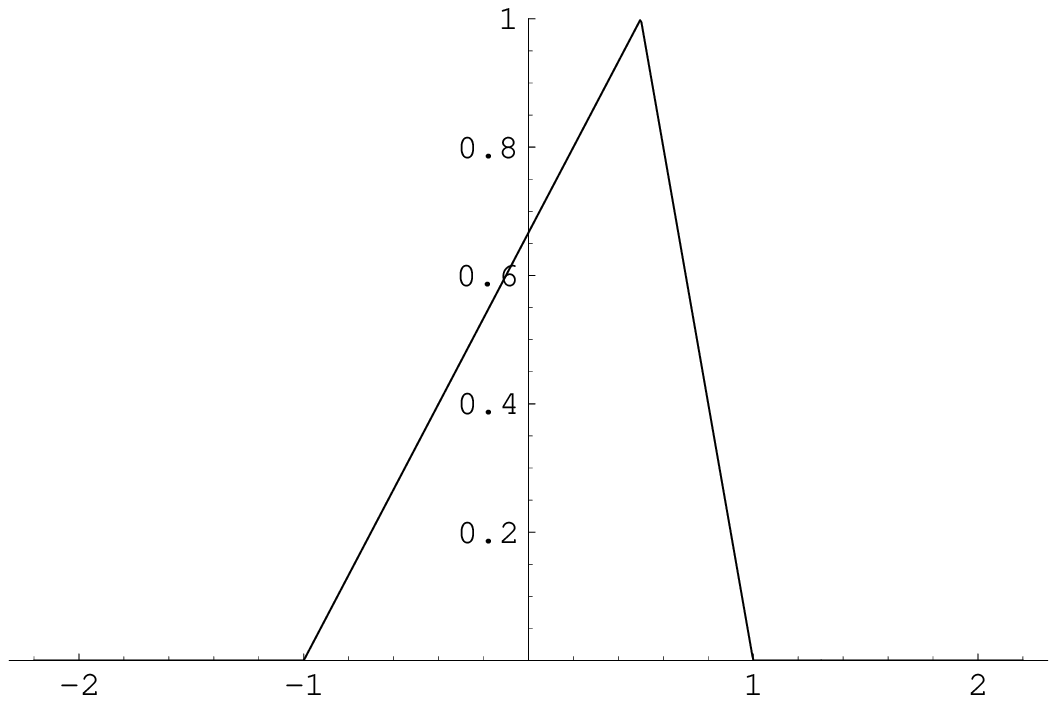,width=0.18\linewidth,
height=4.2cm,clip=} 
& \vspace{-2cm} {\huge $\oplus$}  \vspace{+2cm}&
 \epsfig{file=triang_asy1.eps,width=0.18\linewidth,
height=4.2cm,clip=} 
&\vspace{-2.cm} {\huge $=$} \vspace{+2cm}& 
\epsfig{file=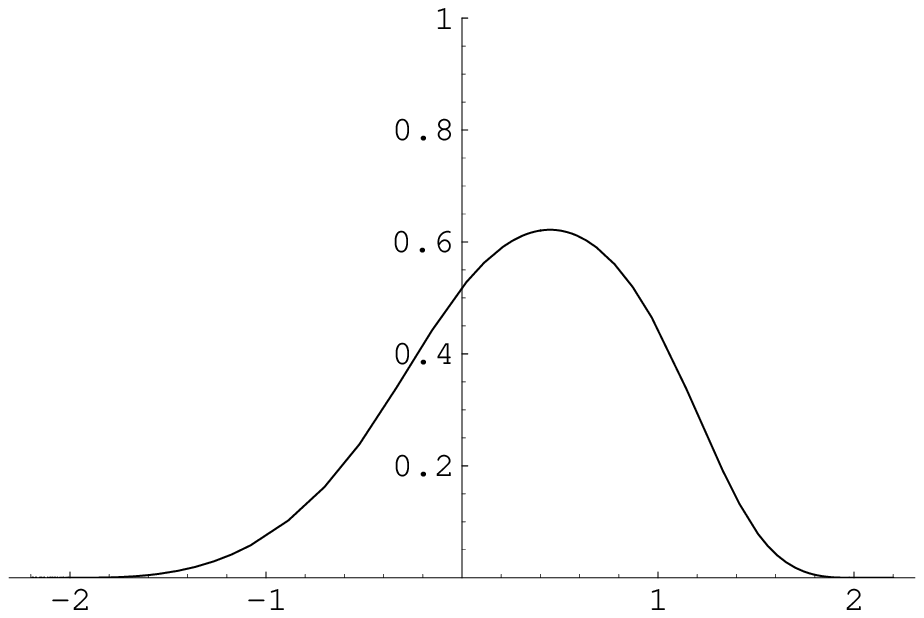,width=0.39\linewidth,clip=} \\
&&&& \\
\hline
\end{tabular}
\end{center}
\caption{\small Probability density function resulting from the
sum of two quantities, each described by an asymmetric
 triangular p.d.f.  with $x_0=0.5$, $\Delta x_+ = 0.5$ and  
 $\Delta x_- = 1.5$. }
\label{fig:2triang}
\end{figure}

\newpage
As a numerical example, let us consider the standard deviations
of input quantities believed to be, with certainty or
with high probability,  in the interval between $-1$ and $+1$.
%Using the models of Fig.~\ref{fig:pdfsys} we have:
\begin{eqnarray}
  \mbox{Uniform:} && 
            \sigma(X) = \frac{1}{\sqrt{3}} \ \,\approx  0.58\,. \\
  \mbox{Symmetric triangular:}  & & 
            \sigma(X) = \frac{1}{\sqrt{6}} \ \,\approx 0.41\,. \\
  \mbox{Asymmetric triangular peaked at 1/2:} 
                && 
            \sigma(X) = \sqrt{\frac{13}{72}} \approx 0.42\,. \\  
  \mbox{Gaussian, 90\% probability interval:} 
                && 
            \sigma(X) =  \frac{1}{1.64}  \, \approx 0.61\,. \\  
   \mbox{Gaussian, 95\% probability interval:} 
                && 
            \sigma(X) =  \frac{1}{1.96}  \, \approx 0.51\,. \hspace{2.5cm}
\end{eqnarray}
We see that, for practical purposes, the differences between the 
$\sigma$'s are irrelevant. Nevertheless, in order to avoid a bias of the 
overall uncertainty, one should try to model each component 
according to the best knowledge of the
physics case, rather than by choosing systematically the 
model which gives the most conservative 
uncertainty.\footnote{The so called coherent bet, a recognised
normative tool to assess probability,  can help a lot to elicitate 
model and parameters of type-B uncertainty. For a 
concise introduction on its concept see Ref.~\cite{ajp}. 
An extensive discussion about its role to force people 
to assess uncertainty in non-standardised situations, 
see Ref.~\cite{priors}.} 
Note that in the case of asymmetric triangular distribution, the 
expected value of $X$ is neither the centre of the interval, nor
the peak of the distribution. In this case we have 
$\mbox{E}[X]=1/6\approx 0.17$. If one uses, incorrectly,  the 
peak value, one introduces a bias which is  
$\approx 80\%$ of a standard deviation. 
As an example, 
Fig.~\ref{fig:2triang} shows the resulting uncertainty 
on the quantity $Y=X_1+X_2$, where the $X_i$ are independent
and their uncertainty is described by identical asymmetrical triangular
distributions.
 The combined result is obtained 
analytically using 
\newpage
\noindent
Eq.~(\ref{eq:prop_delta}).
 One can see how good 
the Gaussian approximation already is and how biased
a result can be, if the best estimate of the sum 
is performed using 
mode or median, and if the final uncertainty 
is evaluated with ad hoc rules of the kind shown in the introduction.

\section{Small deviations from linearity}\label{sec:quadratic}
\begin{figure}
\begin{center}
\begin{tabular}{|c|c|} \hline
\multicolumn{2}{|c|}
{\epsfig{file=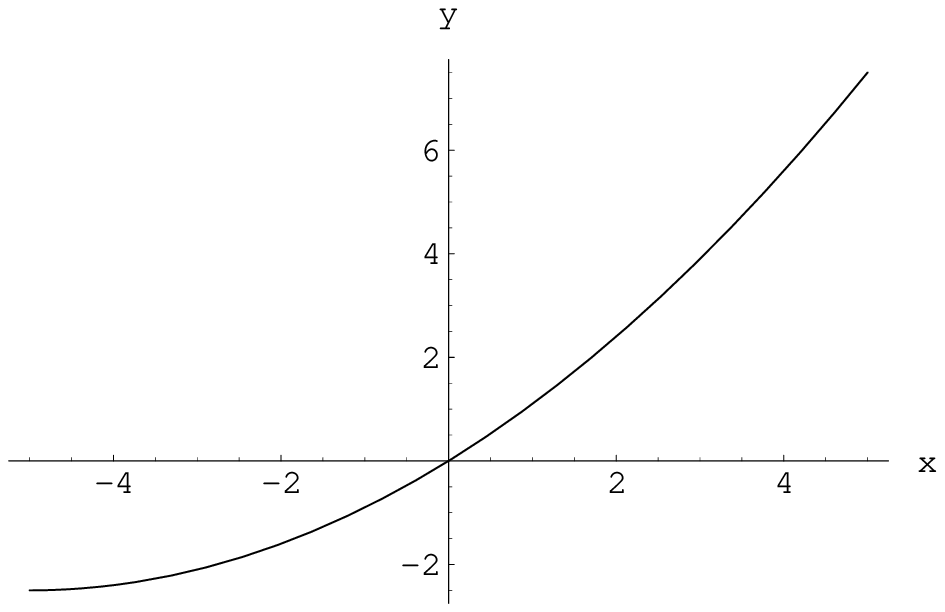,width=0.45\linewidth,clip=}}
\\ \hline
\epsfig{file=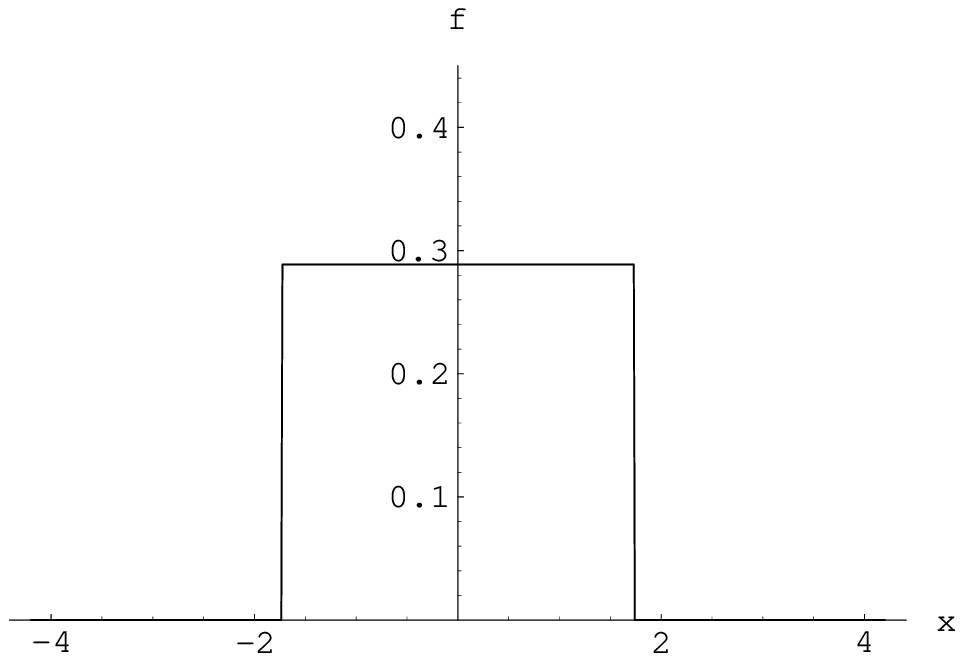,width=0.45\linewidth,clip=} &
\epsfig{file=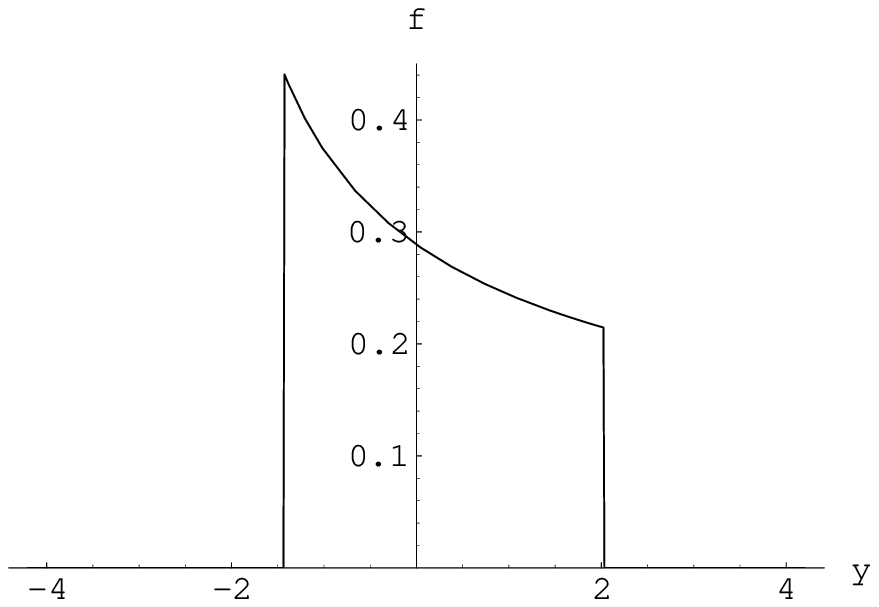,width=0.45\linewidth,clip=} \\ \hline
\epsfig{file=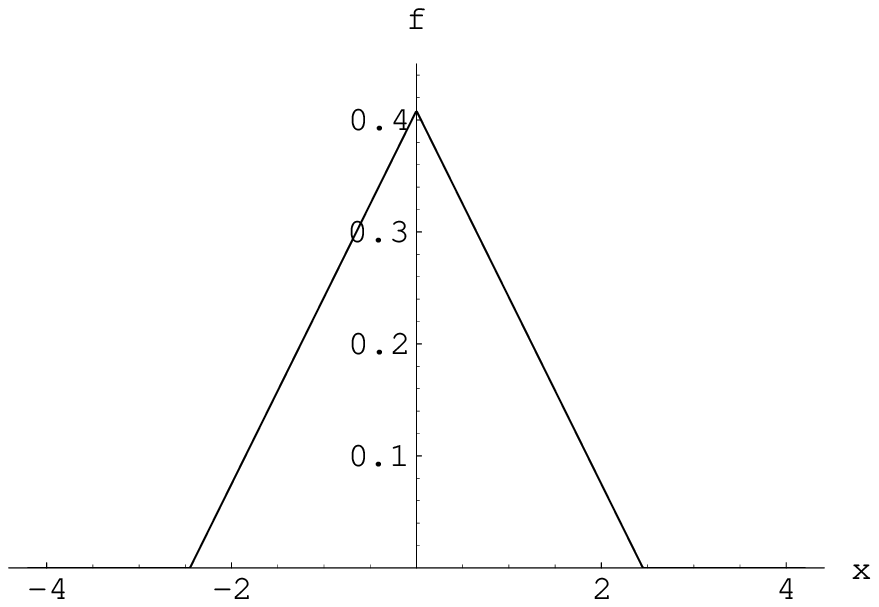,width=0.45\linewidth,clip=} &
\epsfig{file=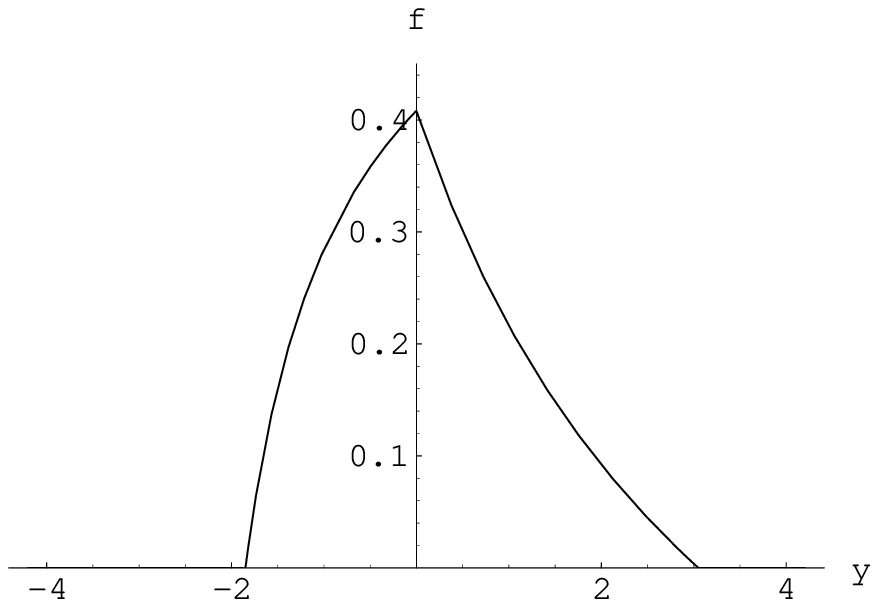,width=0.45\linewidth,clip=} \\ \hline
\epsfig{file=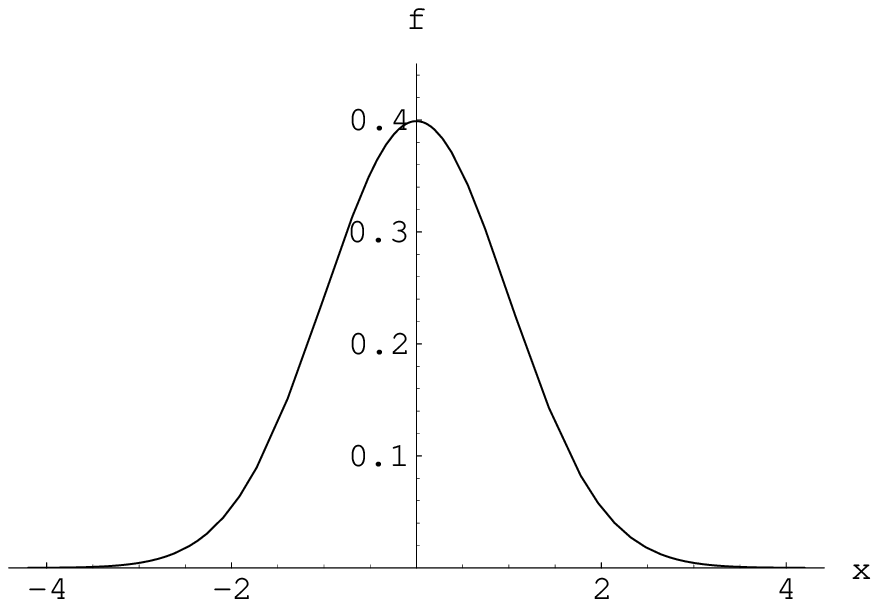,width=0.45\linewidth,clip=} &
\epsfig{file=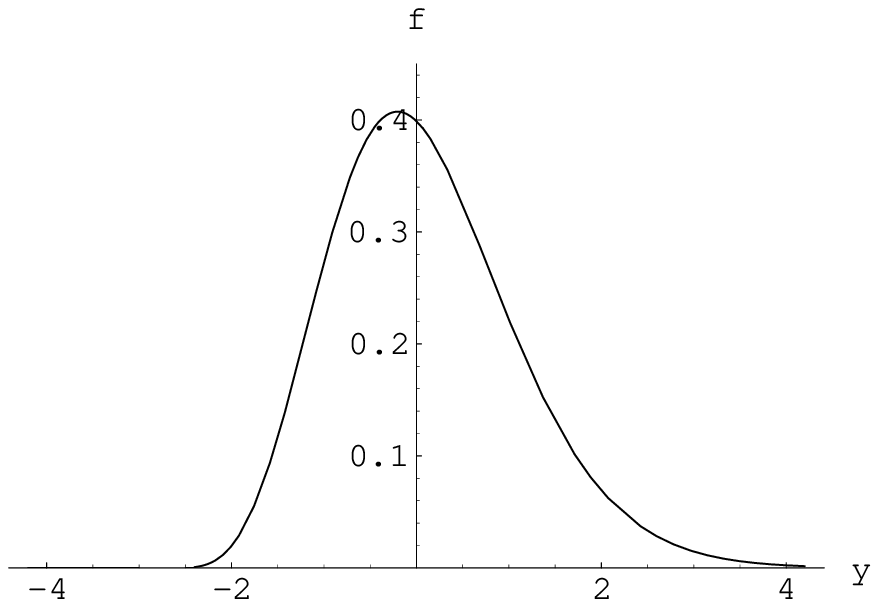,width=0.45\linewidth,clip=}  \\ \hline
\end{tabular}
\end{center}
\vspace{-18.9cm} 
\hspace{+4.0cm}
$Y = \frac{1}{10}X^2+X $ 

\vspace{+4.cm} \hspace{+1.2cm} $X_1$ \hspace{7.0cm} $Y_1$

\vspace{+4.5cm} \hspace{+1.2cm} $X_2$ \hspace{7.0cm} $Y_2$

\vspace{+4.7cm} \hspace{+1.2cm} $X_3$ \hspace{7.0cm} $Y_3$

\vspace{+3.9cm}
\caption{\small Propagation of a uniform, a triangular and a 
Gaussian distribution under a nonlinear transformation. 
The p.d.f.'s of $X_i$ have been evaluated analytically using 
Eq.~(\ref{eq:prop_delta}).}  
\label{fig:quad}
\end{figure}

\begin{figure}
\begin{center}
\begin{tabular}{|c|c|}\hline
\epsfig{file=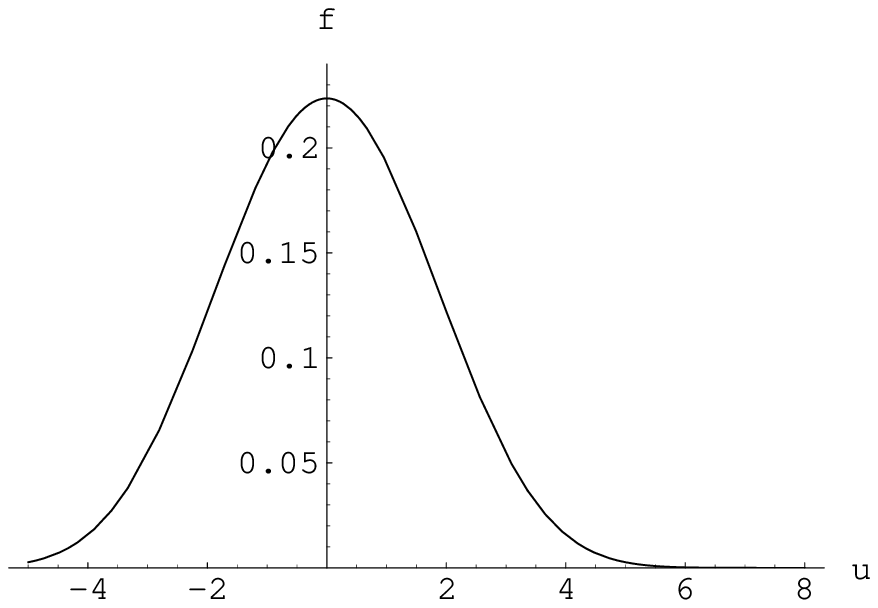,width=0.45\linewidth,clip=} & 
\epsfig{file=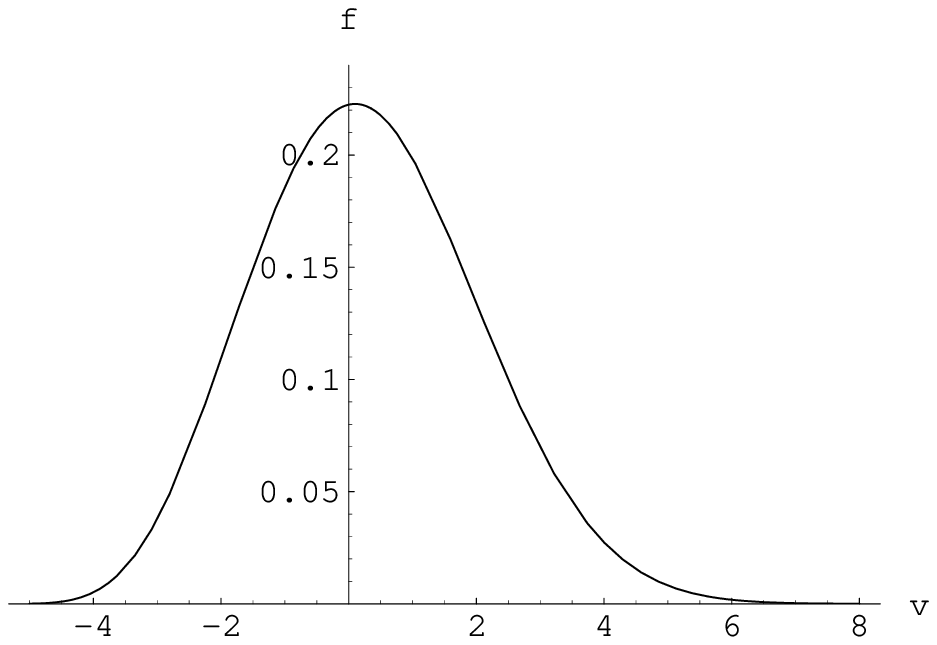,width=0.45\linewidth,clip=} \\ \hline
\end{tabular}
\label{fig:sum_quad}
\end{center}
\vspace{-4.8cm}\hspace{4.0cm}$U=X_1+X_2+X_3$ 
               \hspace{4.3cm}$V=Y_1+Y_2+Y_3$ 
\vspace{+4.3cm}
\caption{\small Probability density functions of the
sum of the quantities $X_i$ and of their nonlinear
transformations $Y_i$ defined in  Fig.~\ref{fig:quad}.} 
\label{fig:quad2}
\end{figure}

\vspace{0.2cm}
\noindent
Let us consider now nonlinearity effects, which are 
mostly responsible for the published asymmetric uncertainties
due to systematics.
Nonlinearity makes in fact $\Delta _{+_{ji}}$ 
and $\Delta _{-_{ji}}$ differ considerably. 
We treat here only second-order effects.
Figure \ref{fig:quad} shows an example of the transformation of 
some important p.d.f.'s, all characterized by $\mbox{E}[X]=0$
and $\sigma(X)=1$, while Fig. \ref{fig:quad2} shows the convolution of 
the original and of the transformed quantities 
of Fig. \ref{fig:quad}. 
One can see that indeed the p.d.f. of the sum of 
of both the original and the transformed quantities 
can be described by a Gaussian for the practical purposes 
of interest in uncertainty evaluations. 

In order to simplify the formulae, let us consider first
the case of only one input quantity and one output quantity
(see Appendix B for the general case).
Taking the second-order expansion, we have
\begin{equation}
Y = Y(\mbox{E}[X]) +
    \frac{\partial Y}
         {\partial X}\,(X-\mbox{E}[X]) +
    \frac{1}{2}\frac{\partial^2 Y}
                    {\partial X^2}\,(X-\mbox{E}[X])^2\,.
\end{equation}
Expected value and variance of $Y$ are then
\begin{eqnarray}
\mbox{E}[Y] &=& Y(\mbox{E}[X]) +
              \frac{1}{2}\,\frac{\partial^2 Y}
                                {\partial X^2}\,\sigma^2(X)\,,
\label{eq:EYnl}\\
\sigma^2(Y) &=&
             \left( \frac{\partial Y}{\partial X}\right)^2
             \sigma^2(X) + 
            \frac{\partial Y}{\partial X}\,
                \frac{\partial^2 Y}
                     {\partial X^2}\,
                    \mbox{E}\left[(X- \mbox{E}[X])^3\right] \nonumber\\
    && \! + \   \frac{1}{4}\, \left( 
               \frac{\partial^2 Y}{\partial X^2} \right)^2
           \left[\mbox{E}[(X-\mbox{E}[X])^4]-\sigma^4(X)\right]\,.
\label{eq:varYnl} 
\end{eqnarray}
These formulae can be transformed into more practical ones if the
derivatives are replaced by their numerical evaluations from the
$\pm 1\,\sigma$ of $X$ around $\mbox{E}[X]$, which produce
variations $\Delta _{\pm}$ in $Y$. The approximate derivatives 
evaluated in $E[X]$ are
\begin{eqnarray}
\frac{\partial Y}{\partial X} &\approx& \frac{1}{2}
       \left(\frac{\Delta _+}{\sigma(X)}+
\frac{\Delta _-}{\sigma(X)}\right) =
       \frac{\Delta _+ +\Delta _-}{2\,\sigma(X)}\,, \\
\frac{\partial^2 Y}{\partial X^2} &\approx &
       \frac{1}{\sigma(X)}\,
       \left(\frac{\Delta _+}{\sigma(X)}-\frac{\Delta _-}{\sigma(X)}\right)
       = \frac{\Delta _+-\Delta _-}{\sigma^2(X)}\,.
\end{eqnarray}
The formula of the variance, Eq.~(\ref{eq:varYnl}),
can be simplified using 
skewness (${\cal S}$) and kurtosis (${\cal K}$), defined as 
${\cal S}=\mbox{E}\left[\left(X-\mbox{E}[X]\right)^3\right]/\sigma^3(X)$
and ${\cal K}=\mbox{E}\left[\left(X-\mbox{E}[X]\right)^4\right]/\sigma^4(X)$,
respectively. We get finally 
\begin{eqnarray}
\mbox{E}[Y] &=& Y(\mbox{E}[X]) + \delta\,, 
\label{eq:EY_nonlinear} \\
\sigma^2(Y) &=& \overline{\Delta}^2 + 
2\,\overline{\Delta}\cdot\delta\cdot S(x)+
                \delta^2\cdot\left[{\cal K}(X)-1\right]\,,
\label{eq:sig_nonlinear}
\end{eqnarray}
where $\delta$ is the semi-difference of the two shifts 
[$\delta = (\Delta _{+}-\Delta _{-})/2$]  and 
$\overline{\Delta}$ is their average 
[$\overline{\Delta}=(\Delta _{+}+\Delta _{-})/2$]. 
The interpretation of Eq.~(\ref{eq:EY_nonlinear}) is simple 
and corresponds to a procedure that some might have 
already guessed: Asymmetric uncertainties produce a shift 
in the best estimate of the quantities. 
 In the case that the dependence between 
$Y$ and $X$ is linear, $\delta$ is $\approx 0$ and we recover the 
result given in Section \ref{sec:linear}. 
Note also that the second term of Eq.~(\ref{eq:sig_nonlinear})
disappears if the distribution describing the
uncertainty on $X$ is symmetric around $\mbox{E}[X]$, and that 
the third term plays a second-order role, since the difference 
between $\Delta _+$ and $\Delta _-$ is usually smaller than their sum,
 and  ${\cal K}(X)$ is around 2 or 3
for the distributions of interest (see Fig. \ref{fig:pdfsys}). 

The extension to several independent input 
quantities is straightforward,
as one only needs to add together the individual 
contributions to expected value and the variance. 
Considering the most common case in which the second and third terms 
of the r.h.s. of Eq.~(\ref{eq:sig_nonlinear}) are 
negligible\footnote{For symmetric distributions the skewness is zero,
while the kurtosis is around 3 for the distributions of interest
and enters with $\delta^2$.},
we obtain the following simple practical formulae: 
\begin{eqnarray}
\mbox{E}[Y] &\approx& Y(\mbox{E}[{\bf X}]) + \sum_i \delta_i\,, \\
\sigma^2(Y) &\approx& \sum_i \overline{\Delta}^2_i\,.
\label{eq:nl_simple}
\end{eqnarray}
Averaging positive and negative deviations is indeed a good
practice, but the shift of the central value should not
be neglected. For the separation of input quantities into $\mu_{r_i}$
and influence factor, see Eqs.~(\ref{eq:varY2r})--(\ref{eq:covY2r}). 
The formulae for the more general case of 
several output quantities and of
correlations among input quantities will be considered 
in Appendix B. 

\vspace{0.2cm}
\section{Numerical examples}\label{sec:examples}
\vspace{0.2cm}
\noindent
Let us go back to the numerical example of the introduction. 
Those numbers were obtained from a quadratic dependence of $Y$ 
on the influence quantities, each having a slightly 
different functional form and a different model
to describe its uncertainty. 
Including also $\mu_r$ as $X_0$,
we can write the dependence of $Y$ on $X_i$ 
 in the following explicit form: 
\begin{equation}
Y= \sum_{i=0}^{3} \alpha_i X_i + \beta_i \,X^2_i\,,
\end{equation}
where $\alpha_i$ and $\beta_i$ are given in Table \ref{tab:pdfsys2}, 
in which also the uncertainty model is indicated.  
\begin{table}[t]
\caption{\small 
Parameters of the input quantities used in the numerical
example of the text. $X_0$ is identified with 
the value $\mu_r$, obtained when $X_{1-3}$ are equal to 
their expected values.}
\vspace{0.4cm}
{\centering 
\begin{tabular}
{clcccccc}
\hline
\multicolumn{8}{c}{Interpretation 1: `reasonable variations' 
= $\pm 1\,\sigma$ for all $X_i$} \\
\hline 
Input/Output & \multicolumn{1}{c}{Distribution} 
  &E[X] & $\sigma(X)$ & $\alpha$ & 
$\beta$ & $\Delta_-$ & $\Delta_+$ \\
\hline 
$X_0 (\equiv \mu_r)$ &  Gaussian &1 & 0.05&1 & 0 & 
$ + 0.050 $ & $+0.050$ \\ 
$X_1$ & Gaussian & 0 &0.3&0.25 & -0.167 
      & $+0.090$ & $+ 0.060$ \\ 
$X_2$& Triangular  $[-1,1] $ &0 &$0.41$
                              & 0.30 &-0.147  & 
       $+0.147 $ & $+0.098$   \\ 
 $X_3$& Uniform $[-1,1]$& 0 & $0.58$ &  0.225 & -0.078&
      $+0.156$ & $+0.104$ \\ \hline
$Y$ & $\approx$ Gaussian & 0.93 & 0.20 & & & & \\
&&&&&&& \\
\multicolumn{8}{c}{Interpretation 2: `reasonable variations' 
= $\pm 1\,\sigma$ for $\mu_r$ and $X_1$; $\pm\Delta x$ for others}\\
\hline 
Input/Output & \multicolumn{1}{c}{Distribution} 
  &E[X] & $\sigma(X)$ & $\alpha$ & 
$\beta$ & $\Delta_-$ & $\Delta_+$ \\
&&&&&& \multicolumn{2}{c}{(rescaled at 1\,$\sigma$)} \\
\hline 
$X_0 (\equiv \mu_r)$ &  Gaussian &1 & 0.05&1 & 0 & 
$ + 0.050 $ & $+0.050$ \\ 
$X_1$ & Gaussian & 0 &0.3&0.25 & -0.167 
      & $+0.090$ & $+ 0.060$ \\ 
$X_2$& Triangular  $[-1,1] $ &0 &$0.41$
                              & 0.123 &-0.0245  & 
       $+0.054 $ & $+0.046$   \\ 
 $X_3$& Uniform $[-1,1]$& 0 & $0.58$ &  0.130 & -0.026&
      $+0.084$ & $+0.066$ \\ \hline
$Y$ & $\approx$ Gaussian & 0.97 & 0.13  & & & & \\
\hline 
\end{tabular}
}
\label{tab:pdfsys2}
\end{table}
As stated in the introduction, the expression `reasonable variation 
of the parameters' was intentionally left vague.
We consider the two cases in which the variations of non-Gaussian quantities
correspond to $\pm\,1\,\sigma$  or to $\pm$ half-interval, respectively
(`interpretation 1' and `interpretation 2' in Table~\ref{tab:pdfsys2}).
The details of the first evaluation are
(see Appendix B for the second case and for the values 
of central moments of higher order)
\begin{eqnarray}
\mbox{E}[Y] &=& 1.00 + \sum_i \delta_i = 1.00 +
                           (-0.015-0.026-0.0245)=0.9345\,,
\label{eq:res_media}\\
\sigma^2(Y) &=& \sigma^2_r(Y)+ \sigma^2_{sys}(Y)
                   =(0.05)^2+(0.1983)^2= (0.2046)^2\,,
\label{eq:res_sigma}
\end{eqnarray}
a result which can be summarized as $Y=0.93(0.20)$, or
$Y=0.93\pm 0.20$, although the latter expression might 
be misleading, since it is traditionally used as a 68\% 
probability interval, which is exactly true only if the p.d.f. is perfectly
Gaussian\footnote{As stated above and shown 
with several figures, in most of the practical
cases the Gaussian approximation is a good one, even if the 
models describing the uncertainty of the input quantities 
are not Gaussian or if there are some nonlinearity effects. 
In summarizing the result with a couple of numbers, our preference
goes to expected value and standard deviation, because these are
the parameters which matter in further propagation of uncertainty,
and this is the most common use of scientific results. 
In this respect we agree with the recommendations of the
ISO Guide~\cite{ISO}. 
In the case that the final p.d.f. differs considerably from a Gaussian,
is indeed a good practice to provide \underline{also}
mode and median of the distribution, as well as probability
intervals of interests (see e.g. Refs. \cite{Higgs} and 
\cite{sceptical}). 
However, it is clear that if, for decision
problems, one wishes to assess in the most precise way
probability intervals, the exact form of the p.d.f. is required. 
For complex problems this evaluation is performed by 
numerical or Monte Carlo techniques. 
We give in Appendix B approximate formulae of 
skewness and kurtosis, which give an idea of the deviation of 
the p.d.f. from a Gaussian. Appendix C shows also how to 
get an idea of the shape of the p.d.f., assuming that it is 
not `too' different from a Gaussian.   
} 
(see also comments in Ref~\cite{ISO}). The
result 
(\ref{eq:res_media})--(\ref{eq:res_sigma}) is in excellent agreement
with $\mbox{E}[Y] = 0.9344$ and 
$\sigma(Y) = 0.2046$ obtained directly from the p.d.f. of $Y$
estimated by Monte Carlo with $10^6$ extractions.
In contrast, the result obtained combining separately positive and 
negative deviations in quadrature (see introduction) 
shows a bias which amounts to 35\% of $\sigma$. 
%the standard deviation. 

\vspace{0.2cm}
\section{The non-monotonic case}\label{sec:parabolic}
\vspace{0.2cm}
\noindent
Sometimes a variation of $\pm\,1\sigma$ 
of an influence parameter might produce values of $Y$ which are 
both above or both below the value obtained with the 
reference value. Using the notation of this paper, 
$\Delta_+$ and $\Delta_-$ have opposite signs in that case. 
This result indicates that the function is not monotonic,  
and this situation has to be treated with some care. In fact, although the 
formulae derived in this paper do not depend on whether the 
functions are monotonic 
or not, the transformed distribution can be very different 
from those of Fig.~\ref{fig:quad}  and  can bring 
a large non-Gaussian contribution to the overall distribution. 
\begin{figure}[t]
\begin{center}
\begin{tabular}{|c|c|} 
\hline
\epsfig{file=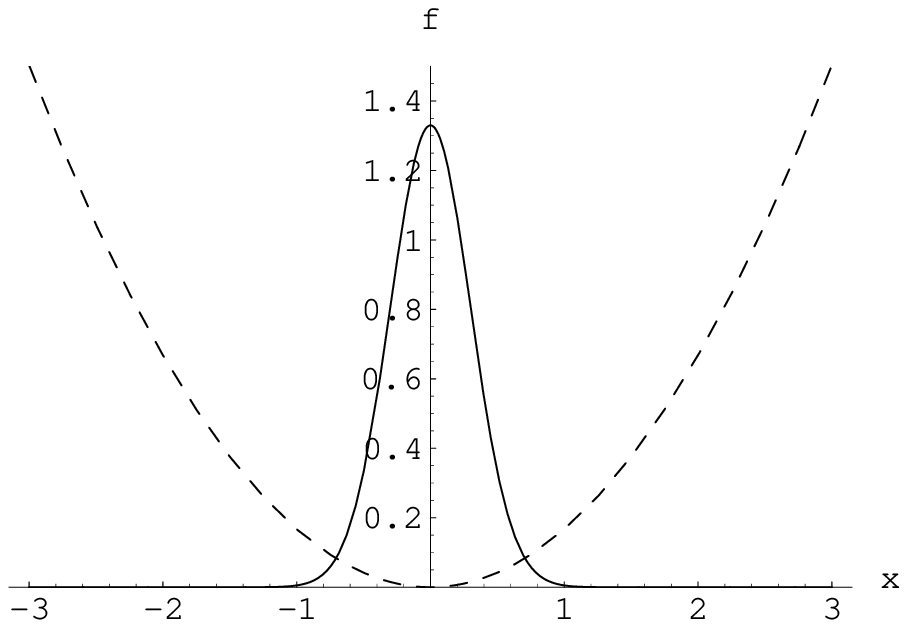,width=0.44\linewidth,clip=} 
 & \epsfig{file=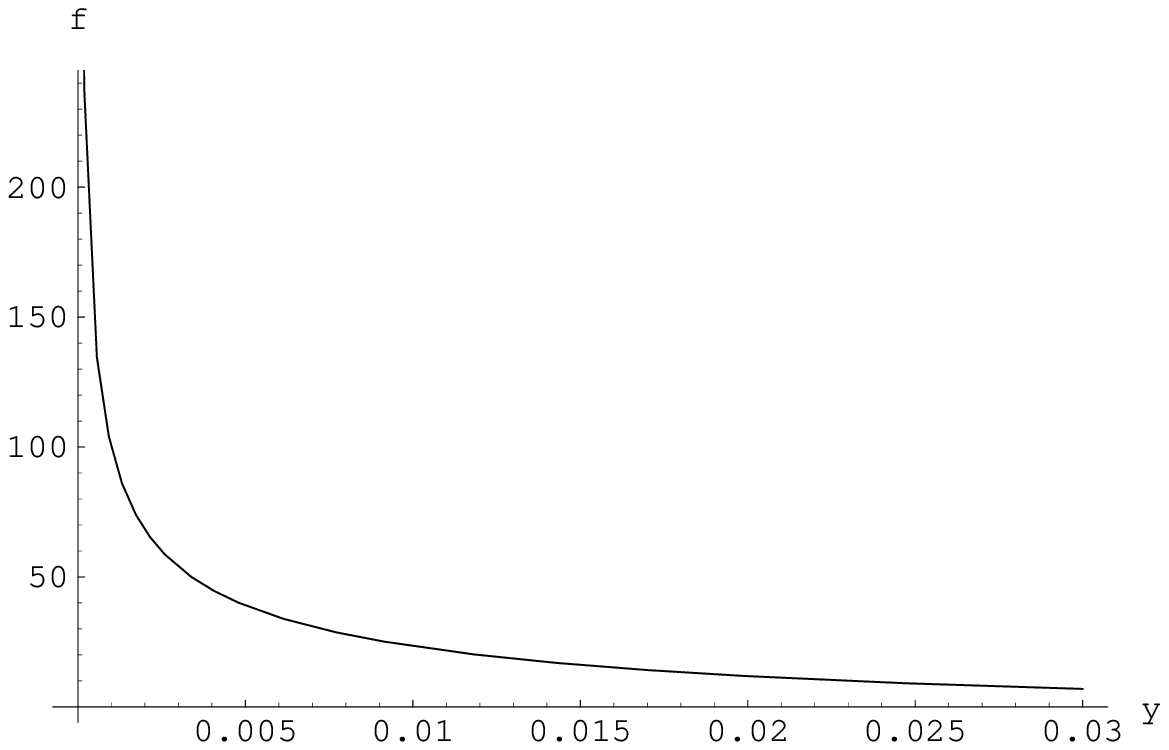,width=0.5\linewidth,clip=} \\ \hline
\end{tabular}
\end{center}
\vspace{-5.4cm}\hspace{0.5cm}$Y=Y(X)$
\vspace{+5.0cm}
\caption{\small Example of non-monotonic relation between 
input and output quantity. The left plot show the 
parabolic dependence of $Y$ on $X$ (dashed line) 
and the Gaussian p.d.f. of $X$
(solid line). The right plot shows the p.d.f. of $Y$.} 
\label{fig:parabola2}
\end{figure}
As an example, let us consider Fig.~\ref{fig:parabola2}, which
describes an input quantity normally distributed around 0
with $\sigma=0.3$, 
a parabolic dependence of $Y$ on $X$ given by 
$Y=0.167\,X^2$ (i.e. like $X_1$ of Table~\ref{tab:pdfsys2}, but with
the $\alpha=0$ and $\beta$ reversed in sign, 
just for graphical convenience). 
The $\pm 1\,\sigma$ variations are $\Delta_+ = +0.015$ and 
$\Delta_- = -0.015$, but certainly one would not quote 0
as the expected value  of $Y$, nor 0.015 its standard deviation. 
$\mbox{E}[X]$ being at the minimum of the distribution, 
the p.d.f. of X ends sharply at zero, and is very asymmetric. 
In fact it is easy to recognize in $f(y)$ a scale transformation of 
the $\chi^2$ with one degree of 
freedom, namely $Y=0.015\times\chi^2_1$. Expected values 
and standard deviation are then $\mbox{E}[Y]=0.015$ and
$\sigma(Y)=0.015\times\sqrt{2}=0.021$. 
We can compare the result with what we get from 
Eqs.~(\ref{eq:EY_nonlinear})--(\ref{eq:sig_nonlinear}): 
\begin{eqnarray}
\left\{ \begin{array}{rcl} \overline{\Delta} & = & 0 \\
                           \delta            & = & 0.015 
         \end{array} \right.
\ \Longrightarrow\ \ 
\left\{ \begin{array}{rcl}  \mbox{E}[Y]  &=& 0 +  0.015 = 0.015  \\ 
                     \sigma^2(Y)       &=& 0 + 0 + 0.015^2\times 2= (0.021)^2
         \end{array} \right.
%\overline{\Delta} &=& 0\,,     \nonumber\\
%\delta            &=& 0.015\,, \nonumber\\
%E[Y]              &=& 0 +  0.015 = 0.015\,, \nonumber \\
%\sigma^2(Y)       &=& 0 + 0 + 0.015^2\times 2= (0.021)^2\,. \nonumber
\end{eqnarray}
The result is exactly the same, as it should be, since in this
example the function is parabolic and, therefore, there are no
approximations in Eqs.~(\ref{eq:EY_nonlinear})--(\ref{eq:sig_nonlinear}).
We see that in this case only the quadratic terms appear. 
Similarly, it would be wrong to consider the best estimate of 
$Y$ as equal 0, with an uncertainty equal to the deviation: 
The result would have a standard deviation smaller by $\sqrt{2}$, 
and the best estimate would have a bias of $-140\%$ of the reported 
standard deviation. 

\vspace{0.2cm}
\section{Conclusions}\label{sec:conclusions}
\vspace{0.2cm}
\noindent
The issue of uncertainty propagation  has been reviewed 
using the probabilistic approach nowadays called Bayesian. 
Following physicists' intuition,
this approach allows 
probabilistic reasoning to be applied 
to not exactly-known values of 
physical quantities. Therefore, probabilistic statements 
then have  a clear meaning and  can 
be propagated to the values of related quantities 
using probability calculus. In particular, uncertainties
due to systematic effects or theoretical inputs can be 
included in a conceptually easy way. 
 Indeed, there is no qualitative distinction between uncertainty 
due to random and systematic (or theoretical) effects, 
and in this respect we
agree fully with the ISO Guide~\cite{ISO}.

Approximate formulae have been derived to calculate 
the first moments of output p.d.f., taking into account up to
second-order effects in the uncertainty propagation. 
Non-linear effects are responsible, together with the less
frequent  case of asymmetric uncertainties on input quantities, 
of results given with asymmetric uncertainties, 
if uncertainties are evaluated 
using {\it ad hoc} procedures. We have shown that in most cases
of practical interest, a combinatorial effect similar to
the central limit theorem  symmetrizes 
automatically the overall uncertainty. It follows that 
expected value and standard deviation of the distribution
should be the correct way of presenting the result, as also 
pointed out by metrological authorities~\cite{ISO}. 
In the case that the final distribution is considerably non-Gaussian,
other position and width indicators of the distribution 
(and whenever possible the p.d.f. itself) should also be provided, 
but expected value and standard deviation should always be given,
as they are what mostly matters in further propagation 
of uncertainty. 
We have also shown that asymmetric uncertainty on input quantities
and/or non-linear effects produces shifts in the expected value 
of the output quantity with respect to the value calculated 
for the best estimate of the input quantities. {\it If these shifts 
are not applied, the result can be biased.} 

As far as modelling the uncertainty due to systematic effects 
is concerned, we have 
found particularly helpful the concept of type-B uncertainty,
according to the ISO terminology~\cite{ISO}. We have shown that 
the results are rather stable against reasonable variations
of models and values of the parameters. On the other hand,
sticking to the position that 
there is no way of modelling uncertainty due to systematic effects
would lead to impossibility of providing an overall uncertainty, 
or to providing this uncertainty 
using arbitrary prescriptions which do not have 
a clear meaning. In other words,
we prefer beliefs assessed  by people we trust, rather 
than empty prescriptions, in accordance with the scheme: beliefs in, 
beliefs out; nothing in, nothing out (see Ref.~\cite{ajp}
for a discussion about `belief' and `arbitrariness'). This 
position is based on the assumption that 
{\it ``it is scientific
only to say what is more likely and what it is 
less likely''}~\cite{Feynman}.

We would like to conclude with some remarks about when to stop
making systematic checks and adding contributions 
to the overall uncertainty. As eloquently said~\cite{Wahl}, 
{\it ``one could correlate
the result with the phase of the Moon or the position of Jupiter,  
and find most likely no significant effect,
with some uncertainty; but certainly we don't want to
take care of this uncertainty.''} Only contributions
which are in principle relevant should be considered 
in the uncertainty evaluation. Even if the effect 
is `statistically significant', one should try to understand if
it can physically influence the result, before including it
in the analysis. There is certainly some 
subjectivity in deciding what is relevant and what is not, 
but this is consistent with the spirit that {\it ``the 
quality and the utility of the uncertainty quoted for the 
result of a measurement therefore ultimately depend on the 
understanding, critical analysis, and integrity of those 
who contribute to the assignment of its value''}\cite{ISO}.

\newpage
\section*{Appendix A -- Derivation of Eq.~(\ref{eq:prop_delta})
and comparison with Eq.~(\ref{eq:inf_cond2})}
\vspace{0.2cm}
\noindent
Equation (\ref{eq:prop_delta}) is one of the most convenient ways 
to formulate the problem of p.d.f. propagation, 
especially for physicists, who are familiar with the Dirac delta. 
Better known formulae which make use of the Jacobian 
are easily recovered by making use the properties of the delta. 
The same is true for the convolution formula to obtain 
the p.d.f. of the sum of two independent quantities. 
In this latter case, the derivation is straightforward: 
$f(y)=\int f(x_1)\cdot f(x_2)\cdot \delta(y-x_1-x_2)\,
      \mbox{d}x_1\mbox{d}x_2 = 
      \int f(x_1)\cdot f(y-x_1)\,\mbox{d}x_1$.  
Equation (\ref{eq:prop_delta}) is also important because it justifies
the Monte Carlo estimation of the p.d.f. and clarifies the 
role of the various ingredients which enter the game 
(see comments in footnote 7). 

Although we think that Eq.~(\ref{eq:prop_delta}) does not 
need to be proved, as it can be considered itself the formulation
of the problem, we give here a formal derivation based 
on the properties of the characteristic function. 
%Let $X_i$ be the set of input quantities, on which
%$Y$ depends: $Y=g(X)$. 
The  characteristic function associated with $X$ 
is defined as (see e.g. \cite{kendall}): 
\begin{equation}
\phi_X (t)\equiv \mbox{E}\left[e^{itX}\right] 
= \int\! e^{itx} f(x)\,\mbox{d}x\,,
\end{equation}
from which the p.d.f. can be reobtained as 
\begin{equation}
f(x) = \frac{1}{2\pi}\int \! e^{-itx} \phi_X (t)\,\mbox{d}t.
\end{equation}
If we have a function $Y=g(X)$, the p.d.f. of $Y$ 
can be obtained from the following 
 characteristic function~\cite{kendall}:
\begin{equation}
\phi_Y (t) = \int\! e^{itg(x)} f(x)\,\mbox{d}x\,.
\end{equation}
This property can be extended to a variable $Y$ depending 
on many variables, i.e.
\begin{equation}
\phi_Y (t)= \int\! e^{itg({\bf x})} f({\bf x})\,\mbox{d}{\bf x}\,.
\end{equation}
It follows
\begin{equation}
f(y) = \frac{1}{2\pi}\int \! e^{-ity} \phi_Y (t)\,\mbox{d}t= 
\int\! e^{-ity}\,\mbox{d}t \int\! e^{it g(x) } f({\bf x})\,\mbox{d}{\bf x} = 
\int \! e^{-it (y - g(x)) } f({\bf x})\, \mbox{d}{\bf x} \mbox{d}t .
\label{eq:anti_char}
\end{equation}
Noting that $\int \! e^{it(x-y)}\,\mbox{d}t = 2\pi\,\delta(x-y)$
we get finally 
\begin{equation}
f(y) = \int\! f({\bf x})\,\delta(y-g({\bf x})) \, \mbox{d}{\bf x}\,,
\end{equation}
which is equivalent to Eq.~(\ref{eq:prop_delta}).

Once we have proved Eq.~(\ref{eq:prop_delta}), there is
no need to prove also that this is equivalent to 
 Eq.~(\ref{eq:inf_cond2}), as the latter comes from a general
theorem of probability theory. 
We show their equivalence in a simple case,
using the example of the offset 
uncertainty given in Section \ref{sec:prob}. Using the notation
of Eq.~(\ref{eq:prop_delta}), the correspondence of symbols is
$Y\equiv \mu$,
 $h_1 \equiv z$ (with $z_\circ=0$) and
$\mbox{``data''}\equiv x$. The p.d.f.'s of our input quantities are
\begin{eqnarray}
f(\mu_r\,|\,x) \left[\equiv f(\mu\,|\,x,z_\circ)\right] &=& 
 \frac{1}{\sqrt{2\,\pi}\,\sigma_r}\,
e^{-\frac{(\mu_r-x)^2}{2\,\sigma_r^2}}\,, \\
 & &\nonumber \\
f(z) &=& \frac{1}{\sqrt{2\,\pi}\,\sigma_z}\,
e^{-\frac{z^2}{2\,\sigma_z^2}}\,, 
\end{eqnarray}
while the function that relates output to input quantities is 
$\mu =   \mu_r + z$\,.
Applying Eq.~(\ref{eq:prop_delta}), and writing explicitly 
the integration limits, we have
\begin{eqnarray}
f(\mu\,|\,x) &=& 
%\int_{-\infty}^{+\infty}\! \int_{-\infty}^{+\infty}\! 
%                 f(\mu_r\,|\,x) \cdot f(z)\cdot
%                 \delta (\mu-\mu_r-z)\,\mbox{d}\mu_r\mbox{d}z \\
% & &\nonumber \\
%             &=& 
\int_{-\infty}^{+\infty}\!
\int_{-\infty}^{+\infty}\!
 \frac{1}{\sqrt{2\,\pi}\,\sigma_r}\,
e^{-\frac{(\mu_r-x)^2}{2\,\sigma_r^2}}\,
 \frac{1}{\sqrt{2\,\pi}\,\sigma_z}\,
e^{-\frac{z^2}{2\,\sigma_z^2}}\,\delta(\mu-\mu_r-z)
\,\mbox{d}\mu_r\mbox{d}z \\
 & &\nonumber \\
&=& \int_{-\infty}^{+\infty}\!
 \frac{1}{\sqrt{2\,\pi}\,\sigma_r}\,
e^{-\frac{(\mu-z-x)^2}{2\,\sigma_r^2}}\,
 \frac{1}{\sqrt{2\,\pi}\,\sigma_z}\,
e^{-\frac{z^2}{2\,\sigma_z^2}}
\,\mbox{d}z\,, 
\end{eqnarray}
which is exactly Eq.~(\ref{eq:intz}).

\vspace{1.0cm}
\section*{Appendix B -- Useful formulae up to second-order approximation}
\vspace{0.2cm}
\noindent
In the general case of several input quantities and several 
output quantities the second-order expansion is given
by 
\begin{eqnarray}
Y_j &=& Y_j(\mbox{E}[{\bf X}]) +
            \sum_i \frac{\partial Y_j}
         {\partial X_i}\,(X_i-\mbox{E}[X_i]) \nonumber \\
 && \! +\    \frac{1}{2} \sum_{l,m} \frac{\partial^2 Y_j}
          {\partial X_l \partial X_m}\,(X_l-\mbox{E}[X_l])\,
          (X_m-\mbox{E}[X_m])\,. 
\end{eqnarray}
Expected value, variances and covariances of $Y_j$ are then
\begin{eqnarray}
\mbox{E}[Y_j] &=& Y_j(\mbox{E}[{\bf X}]) +
              \frac{1}{2}\,\sum_{l,m}\frac{\partial^2 Y_j}
             {\partial X_l \partial X_m}\,\mbox{Cov}(X_l,X_m)\,,
\label{eq:EYi} \\
\sigma^2(Y_j) &=& \sum_{l,m}
             \left ( \frac{\partial Y_j}{\partial X_l}\right ) 
             \left ( \frac{\partial Y_j}{\partial X_m}\right ) 
                   \,\mbox{Cov}(X_l,X_m) \nonumber \\
           && \! +\  \sum_{l,m,n}   \frac{\partial Y_j}
                     {\partial X_l}\,
                \frac{\partial^2 Y_j}
                     {\partial X_m \partial X_n}\,
                    \mbox{E}[\widetilde{X_l} \widetilde{X_m}\widetilde{X_n}
                            ] \nonumber \\ 
    && \! +\             \frac{1}{4}\, \sum_{h,l,m,n}\left( 
               \frac{\partial^2 Y_i}{\partial X_h\partial X_l } \right)
        \left( 
               \frac{\partial^2 Y_i}{\partial X_m \partial X_n } \right)
         \nonumber \\ 
&&  \! \times\ \left(\mbox{E}[\widetilde{X_h} 
                     \widetilde{X_l}\widetilde{X_m}\widetilde{X_n}]- 
            \mbox{Cov}(X_h,X_l)\,\mbox{Cov}(X_m,X_n) 
      \right)  \label{eq:varYi} \\
\mbox{Cov}(Y_j,Y_k)&=& \sum_{l,m}  \frac{\partial Y_j}{\partial X_l} 
                         \frac{\partial Y_k}{\partial X_m} 
                         \,\mbox{Cov}(X_l,X_m)  \nonumber  \\
   && \! + \ \frac{1}{2} 
                   \sum_{l,m,n} \left( \frac{\partial Y_k}{\partial X_l}
                   \frac{\partial ^2 Y_j}{\partial X_m \partial X_n}
                + \frac{\partial Y_j}{\partial X_l}
                  \frac{\partial ^2 Y_k}{\partial X_m \partial X_n }\right) 
                 \mbox{E}\left [\widetilde{X_l} 
                    \widetilde{X_m}\widetilde{X_n}\right ] \nonumber \\ 
   && \! + \ \frac{1}{4}\sum_{h,l,m,n}   
                \left( 
             \frac{\partial ^2 Y_j}{\partial X_l \partial X_h}  \right ) 
                \left( \frac{\partial ^2 Y_k}{\partial X_m \partial X_n} 
                \right )  \nonumber \\
   &&   \! \times\  \left (
                   \mbox{E} \left [ 
\widetilde{X_h} \widetilde{X_l}\widetilde{X_m}\widetilde{X_n}
                    \right ]  
              -     \mbox{Cov}(X_h,X_l)\,\mbox{Cov}(X_m,X_n) 
                \right ),  
\label{eq:covYi}
\end{eqnarray}
where $\widetilde{X_i} = X_i-\mbox[X_i]$. The expressions 
involving expected values of products of more than two
 $\widetilde{X_i}$'s can be 
evaluated by iterating the procedure described here, 
as they are functions of $X_i$.
An interesting case for the applications is, using the same notation
as at the end of Section \ref{sec:linear}, that in which we have $N$
input quantities of the kind $\mu_{r_i}$,
and all the remaining ones have the meaning of uncorrelated 
influence quantities. Using the derivatives evaluated numerically
and the definition of skewness and kurtosis,  
we can write the previous formulae as 
\begin{eqnarray}
\sigma^2(Y_j) &=& \sigma_{r_j}^2+\sum_{i>N}\overline{\Delta}_{ji}^2 
\nonumber \\
&& \! + \  
2\,\sum_{i>N}\overline{\Delta}_{ji}\cdot\delta_{ji}\cdot S(X_i)+
    \sum_{i>N}\delta_{ji}^2\cdot\left[{\cal K}(X_i)-1\right]\,, \\
\mbox{Cov}(Y_j,Y_k) &=& \mbox{Cov}(\mu_{r_j},\mu_{r_k})
+ \sum_{i>N} \overline{\Delta}_{ji} \overline{\Delta}_{ki} \nonumber \\
 && \! +\     \sum_{i>N}(\overline{\Delta}_{ji}\,\delta_{ki}+
              \overline{\Delta}_{ki}\,\delta_{ji})\,{\cal S}(X_i) +
   \sum_{i>N}\delta_{ji} \delta_{ki} 
  \left[{\cal K}(X_i) -1 \right]\,,
\end{eqnarray}
where $\overline{\Delta}_{ji} = (\Delta_{+ji}+\Delta_{-ji})/2$ and
 $\delta_{ji} = (\Delta_{+ji}-\Delta_{-ji})/2$
and $\Delta_{\pm ji}$ are defined according to the convention 
of footnote \ref{fn:delta}. 
As explained in the text, 
the terms depending on skewness and kurtosis are negligible in 
most practical cases.   

It might be useful to have also the approximate expressions of
skewness and kurtosis of the output quantities 
$Y_j$, to have a rough idea of how much their p.d.f.'s
differ from the Gaussian. Since the formulae become quite 
awful, we only consider the most common case of 
uncorrelated input quantities. 
\begin{eqnarray}
{\cal S}(Y_j) 
&=& \left[ 3 \sum_i \delta_{ji}\,  \overline{\Delta}_{ji}^2 
   \cdot\left[ {\cal K}(X_i) -1 \right] +
\sum_i \delta_{ji}^3\cdot 
  \left[ {\cal E} (X_i) -3 \, {\cal K} (X_i) +2 \right] 
\right.\nonumber \\
&& \!\! +\, \left.\left.
\sum_i \overline{\Delta}_{ji}^3 \, {\cal S}(X_i)
+ 6 \sum_i \delta_{ji}^2\, \overline{\Delta}_{ji}^2 \cdot
   \left[ {\cal P}(X_i) - 2\, {\cal S}(X_i) \right]
 \right]\right/\sigma^3(Y_j)\,,  \\ \label{eq:skew} 
 {\cal K}(Y_j) &=& 
\left[\sum_i  {\cal K} (X_i) \, \overline{\Delta}_{ji}^4 
+ 6 \sum_{l > m} \overline{\Delta}_{jl}^2 \, \overline{\Delta}_{jm}^2 
+ \,6 \sum_{l\ne m} \overline{\Delta}_{jl}^2\, \delta_{jm}^2 
    \cdot\left[{\cal K} (X_m)-1\right] 
 \right. \nonumber \\
&&\!\!+ \,\,6 \sum_i    \overline{\Delta}_{ji}^2\, \delta_{ji}^2 
  \cdot\left[{\cal E} (X_i) -2\,{\cal K} (X_i)+1\right] 
+6 \sum_{l > m} \delta_{jl}^2\, \delta_{jm}^2\, \cdot
\left[{\cal K} (X_l)-1\right]\cdot
  \left[{\cal K} (X_m)-1\right] 
\nonumber \\
&&\!\! + \,\sum_i \delta^4_i \cdot\left[ {\cal O}(X_i) + 6\,{\cal K} (X_i)-
  4 \,{\cal E} (X_i)-3  \right]
 + 4 \sum_i \delta_{ji}\, \overline{\Delta}_{ji}^3 
\cdot\left[{\cal P} (X_i)-{\cal S}(X_i)\right]
\nonumber  \\ 
&&\!\! +\,\,  4 \,\sum_i \overline{\Delta}_{ji}\, \delta_{ji}^3\, 
  \cdot\left[{\cal H} (X_i)-3\,{\cal P} (X_i)+3\,{\cal S}(X_i)
\right] \nonumber \\
&&\!\! + \left.\left. 24 
   \sum_{l>m} \overline{\Delta}_{jl}\, \overline{\Delta}_{jm} 
        \,\delta_{jl}\, \delta_{jm}\, {\cal S}(X_l)\, {\cal S}(X_m)\right]
  \right/\sigma^4(Y_j) \,,  \label{eq:kurto}  
\end{eqnarray}
where  
$$
\begin{array}{rclcrcl}
{\cal P}(X) &=&\mbox{E}\left [ (X-\mbox{E}[X])^5 \right ]\left/ 
                     \sigma^5\right. \,,& \hspace{0.2cm} & 
{\cal E}(X) &=&\mbox{E}\left [ (X-\mbox{E}[X])^6 \right ]\left/ 
                   \sigma^6 \right.\,, \nonumber \\
&&&&&& \nonumber \\
{\cal H}(X) &=& \mbox{E}\left [ (X-\mbox{E}[X])^7 \right ]\left/ 
                     \sigma^7\right.\,,  & \hspace{0.2cm} &
{\cal O}(X) &=& \mbox{E}\left [ (X-\mbox{E}[X])^8 \right ]\left/ 
                     \sigma^8\right. 
\end{array}
$$
%\begin{eqnarray}
%{\cal P}(X) &=& \mbox{E}\left [ (X-\mbox{E}[X])^5 \right ]\left/ 
%                     \sigma^5\right.\,,\ \ \ \ \ \  
%{\cal E}(X) = \mbox{E}\left [ (X-\mbox{E}[X])^6 \right ]\left/ 
%                   \sigma^6 \right.\,, \nonumber \\
%{\cal H}(X) &=& \mbox{E}\left [ (X-\mbox{E}[X])^7 \right ]\left/ 
%                     \sigma^7\right.\,,\ \ \ \ \ \ 
%{\cal O}(X) = \mbox{E}\left [ (X-\mbox{E}[X])^8 \right ]\left/ 
%                     \sigma^8\right.\,
\nonumber 
%\end{eqnarray}
are higher order scaled central moments, 
analogues of skewness and kurtosis. 
Table \ref{tab:mom} gives these higher order moments for 
most relevant p.d.f.'s used to assess type-B uncertainty.
\begin{table}[!h]
\caption{\small The rescaled central moments of 
fifth to eighth order 
for the distribution of Fig.~\ref{fig:pdfsys}. The parameters 
$\alpha$, $\beta$ and $\gamma$ are given in 
Eqs.~(\ref{eq:alpha})--(\ref{eq:gamma}).}
\begin{center}
\begin{tabular}{ccccc}
\hline
Distribution &${\cal P}(X)$&${\cal E}(X)$&
              ${\cal H}(X)$&${\cal O}(X)$ \\ \hline
Gaussian & 0 & 15 & 0 & 105 \\ 
Uniform  & 0 & $27/7$ & 0 & 9 \\ 
Triangular & 0 & $54/7$ & 0 & $144/5$ \\
Asymmetric triangular& $\beta/\sigma^5$ &
$2/7\times(31-27\,\alpha)$ &  $\gamma/\sigma^7$ & 
$16/5\times(13-27\,\alpha)$ \\ \hline
\end{tabular}
\end{center}
\label{tab:mom} 
\end{table}
\break\hfill
The moments of the asymmetric triangular distribution
are given in terms of the
following parameters: 
\begin{eqnarray}
\alpha &=& \left.\Delta x_-^2 
\Delta x_+^2(\Delta x_- + \Delta_+)^2\right/
(\Delta x_-^2 +\Delta x_- \Delta_+ + \Delta x_+^2)^3\,,
\label{eq:alpha} \\
\beta &=& 
- (\Delta x_-^5+5/2\,\Delta x_-^4 \Delta x_+ 
+\Delta x_-^3 \Delta x_+^2 - \label{eq:beta} \\ 
 && \hspace{0.5cm}\Delta x_-^2 \Delta x_+^3-5/2\, \Delta_- 
\Delta_+^4 -\Delta_-^5)/425.2\,,  \nonumber  \\
\gamma &=& -(2\,\Delta x_-^7+7\,\Delta x_-^6 \Delta x_+  
+9\, \Delta x_-^5 \Delta x_+^2+5\, \Delta x_-^4 \Delta x_+^3- 
\label{eq:gamma}\\ 
&& \hspace{0.5cm}5\, \Delta x_-^3 \Delta x_+^4 - 9\,\Delta x_-^2 \Delta x_+^5
        -7\, \Delta_- \Delta_+^6 -2\,\Delta_-^7)/2915.0\,.
 \nonumber
\end{eqnarray}
As a numerical example, continuing on  from
Section \ref{sec:examples}, we compare the results 
obtained with the approximate formulae with those 
obtained calculating the moments directly from the 
p.d.f. estimated by Monte Carlo. The overall results
are given in Table \ref{tab:compare}. The agreement 
is well above that needed for practical purposes.
\begin{table}[h]
\caption{\small 
Comparison between the moments evaluated using approximate formulae
and those obtained by Monte Carlo, based on the examples
 of Table \ref{tab:pdfsys2}.}
\vspace{0.4cm}
\newcolumntype{.}{D{.}{.}{-1}}
\begin{center}
%\begin{tabular}{cccccc}
\begin{tabular}{c..c..}
\hline
&\multicolumn{2}{c}{`Interpretation 1'}
&&\multicolumn{2}{c}{`Interpretation 2'} \\ \hline 
&\multicolumn{1}{c}{\ \ Monte Carlo} & 
 \multicolumn{1}{c}{\ Approx. formulae} 
& &\multicolumn{1}{c}{\ \ Monte Carlo} & 
\multicolumn{1}{c}{\ Approx. formulae} 
\\ \hline
E[Y] & 0.9344& 0.9345 &  & 0.9722 & 0.9720 \\ 
$\sigma$(Y)& 0.2046 & 0.2046 && 0.1297 & 0.1295   \\ 
${\cal S}$(Y) & \mbox{$-$}0.370 & \mbox{$-$}0.372 && 
                \mbox{$-$}0.318 & \mbox{$-$}0.321 \\ 
$ {\cal K}$(Y) & 2.857 & 2.859 && 3.076 & 3.082  \\
\hline 
\end{tabular}
\end{center}
\label{tab:compare}
\end{table}

\vspace{1.0cm}
\section*{Appendix C -- Approximate evaluation of the p.d.f. from the 
first four central moments}
\vspace{0.2cm}
\noindent
We have seen that the propagation of uncertainty
can be solved either exactly, at the cost of having 
to compute complicated multidimensional integrals, or by 
using approximate
formulae which only give the first moments 
of the distribution. The probabilistic interpretation
of the latter case is based on the assumption that the final
p.d.f. is approximately Gaussian, as is often so in most
cases of practical interest. We would like to show here an 
intermediate solution to the problem, which allows 
the shape of the p.d.f. of each $Y_i$ to be estimated starting from the 
approximate evaluation of the first four central moments (see
Appendix B), and solving only one integral in only one variable. 

\begin{figure}
\begin{center}
\begin{tabular}{cc} 
%\hline
\epsfig{file=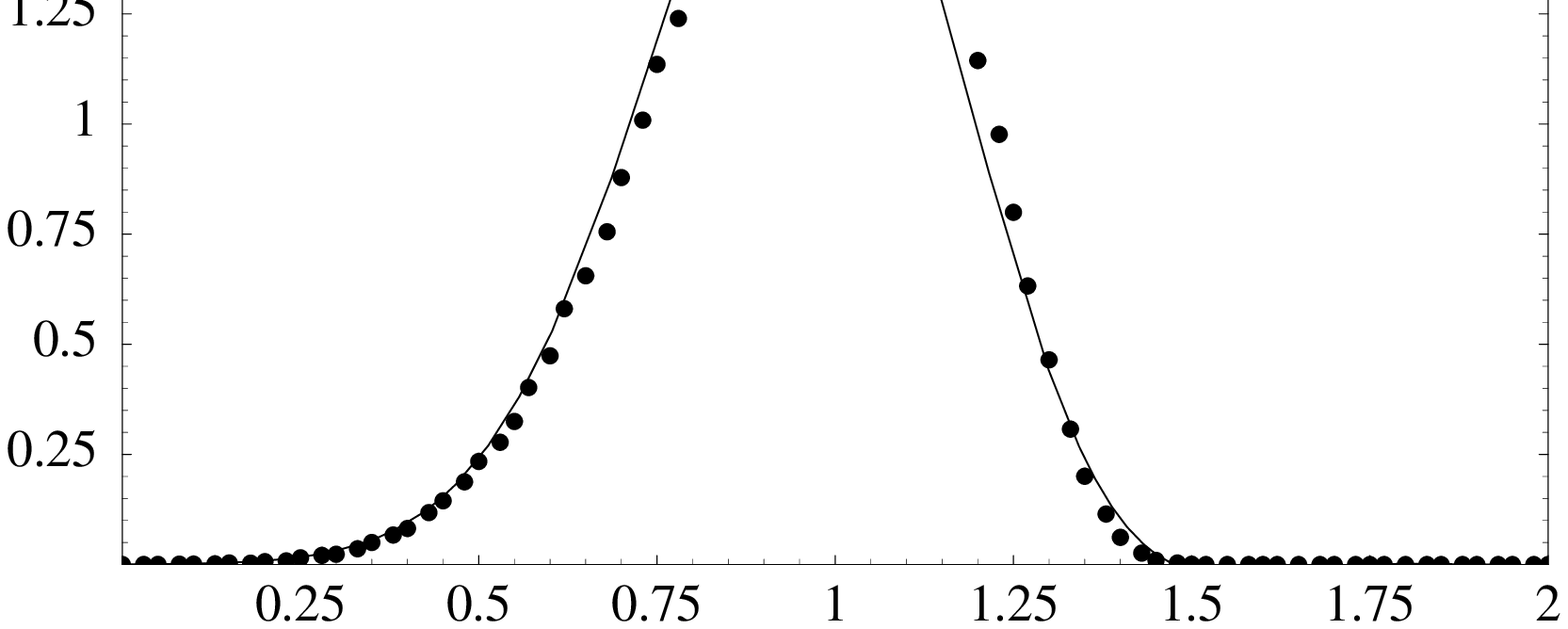,width=0.48\linewidth,clip=} 
 & \epsfig{file=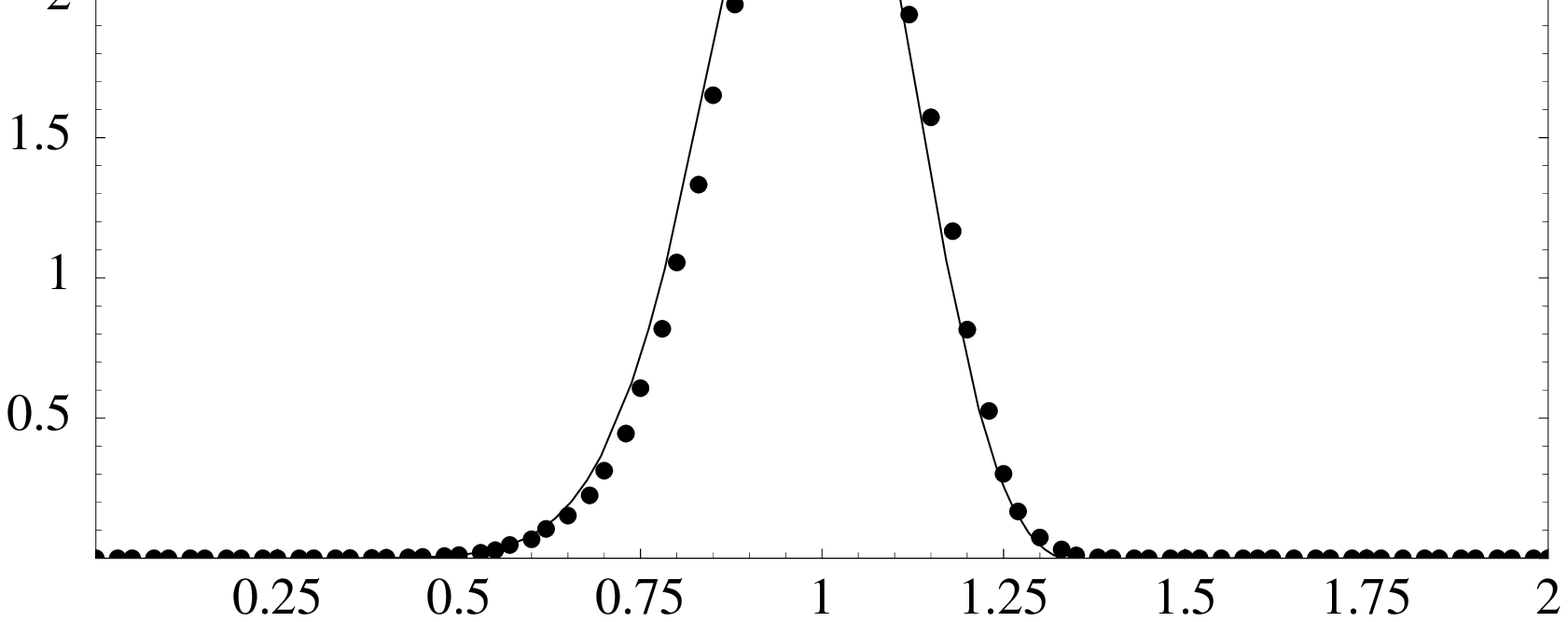,width=0.48\linewidth,height=4.63cm,clip=} 
\end{tabular}
\end{center}
\caption{\small The distribution given by the complete 
uncertainty propagation (dots) is compared with the approximate probability 
density function for interpretations 1 (left side) and 2 (right side).}  
\label{fig:out3}
\end{figure}

Take the generic output variable $Y$.
Its characteristic function $\phi_Y(t)$   
can be expressed as
\begin{equation}
\phi_Y(t) = 
\exp \left ( \sum_{n=1}^{\infty}\frac{\chi_n}{n!}(i\,t)^n  \right ) \,,
%= \exp \left (  i\chi_1t - \frac{1}{2} \chi_2 t^2 - i\frac{1}{3!} \chi_3 t^3 
%   + \ldots  \right )\,,
\end{equation}
where $\chi_n$ is the $n$th `semi-invariant'~\cite{cramer}. 
Semi-invariants can be expressed
in terms of central moments, the first four being
\begin{eqnarray}
 \chi_1 &=& \mbox{E}(X)\,, \nonumber \\ 
 \chi_2 &=& \sigma^2\,, \nonumber \\ 
 \chi_3 &=& \sigma^3\,{\cal S}(X)\,, \nonumber \\
 \chi_4 &=& \sigma^4\,[{\cal K}(X)  -3 ]\,.\nonumber
\end{eqnarray}
An approximate evaluation of the p.d.f. can be obtained under the 
assumption that the first four moments, 
for which we have obtained approximate 
expressions, are the most relevant ones to characterize
the p.d.f.\,. 
This implies that the p.d.f.'s of interest do not differ
 `too much' from a Gaussian. In practice this means that 
they are unimodal, have no discontinuity, and that the 
probability mass is concentrated around a few standard deviations  
from the expected value.  Assuming, then, that all 
 semi-invariants $\chi_n$ with $n>4$ are the same as for a Gaussian 
distribution, namely zero, 
the characteristic function becomes
\begin{equation}
\phi_Y(t) = 
\exp \left ( i\,t\,\mbox{E}[Y] -\frac{1}{2}\,t^2\, \sigma^2(Y) 
-\frac{i}{6}\, t^3\,  \sigma^3\,{\cal S}(Y)
+\frac{1}{24}\,t^4\, \sigma^4 \, [{\cal K}(Y)-3]  \right )\,.
\label{eq:phi_approx}
\end{equation}
%from which we get the p.d.f. of $Y$:
%\begin{equation}
%f(y) = \frac{1}{2\pi} \int\! e^{-ity}\phi_Y(t)\,\mbox{d}t \,.
%\label{eq:pdf_char}
%\end{equation}
%Inserting the expression of $\phi_Y(t)$ in Eq.~(\ref{eq:pdf_char}), 
Inserting Eq.~(\ref{eq:phi_approx}) in the anti-transformation
formula [see Eq.~(\ref{eq:anti_char})] 
and taking only the real part of the solution, 
since the p.d.f. is real,  we get
\begin{equation}
f(y) = \frac{1}{2 \pi} \int \exp\left(-\frac{1}{2}\sigma^2 t^2 +
       \frac{1}{24}\,t^4 \sigma^4\, [{\cal K}(Y)-3]\right)\,  
\cos \left ( t\,[\mbox{E}[Y]-y]-\frac{1}{6} 
t^3\,\sigma^3\,{\cal S}(Y)  \right ) \,\mbox{d}t \, .
\label{eq:integrale}
\end{equation}
Applied to our examples of Section \ref{sec:examples}
and using the values of Table \ref{tab:compare}, 
we get the p.d.f.'s drawn with the continuous curves of 
Fig.~\ref{fig:out3} 
[the integral (\ref{eq:integrale}) has been solved 
with {\it Mathematica}].
For comparison the dotted curves
show the estimations of p.d.f.'s obtained by Monte Carlo. 
The agreement as it can been judged by eye is excellent. 
However, since the method is approximate, 
there are slight problems of normalization and positiveness.
But these problems affect the tails of the distribution
and are not really relevant if one is interested in having an  
idea of the shape under the assumptions of the approximation. 
Note also
the divergent term in Eq.~(\ref{eq:integrale}) for ${\cal K} > 3 $. 
But in practical cases the kurtosis is never 
much larger than this value. In fact one starts usually
from distributions which have ${\cal K}(X) \le 3$ 
(see Fig.~\ref{fig:pdfsys}) and, thanks to the  
central limit theorem, there is a natural tendency 
to have $ {\cal K}(Y) \approx 3$. Therefore, in case of 
values of kurtosis slightly larger that $3$, a good approximation
is to limit it at 3.   
This approximation has been indeed applied to 
the  `interpretation 2' of example of 
Table \ref{tab:compare}, and the resulting p.d.f.
is still in excellent agreement with the Monte Carlo evaluation 
(see right hand plot of  Fig.~\ref{fig:out3}).

\end{document}